\documentclass[12pt]{article}

\usepackage{epsfig,graphicx}
\usepackage{amsmath}
\usepackage{amssymb}
\usepackage{array}

  \newcommand{\query}[1]{\marginpar{%
    \vskip-\baselineskip
    \raggedright\footnotesize
    \itshape\hrule\smallskip#1\par\smallskip\hrule}}
  \newcommand{\removequeries}{\renewcommand{\query}[1]{}}
  
\newcommand{\beq}{\begin{equation}}
\newcommand{\eeq}{\end{equation}}
\newcommand{\bea}{\begin{eqnarray}}
\newcommand{\eea}{\end{eqnarray}}
\newcommand{\no}{\nonumber}

\newcommand{\OMIT}[1]{{}}

\newcommand\spur{\raise.15ex\hbox{/}\kern-.57em }

\newcommand{\cO}{\mathcal{O}}

\newcommand{\CL}{\mathcal{L}}

\newcommand{\upmns}{{U_{\text{PMNS}}}}
\newcommand{\udagpmns}{{U^\dagger_{\text{PMNS}}}}
\newcommand{\ustapmns}{{U^*_{\text{PMNS}}}}

\newcommand{\hc}{\text{h.c.}}
\newcommand{\BR}{\mathcal{B}}
\newcommand{\cB}{\mathcal{B}}
\newcommand{\lsim}{\mathrel{\hbox{\rlap{\hbox{\lower4pt\hbox{$\sim$}}}\hbox{$<$}}}}
\newcommand{\gsim}{\mathrel{\hbox{\rlap{\hbox{\lower4pt\hbox{$\sim$}}}\hbox{$>$}}}}

\def\npb#1#2#3{    {Nucl. Phys.}~B {\bf #1}, #3 (#2)}
\def\plb#1#2#3{    {Phys. Lett.}~B {\bf #1}, #3 (#2)}

\def\prl#1#2#3{    {Phys. Rev. Lett.}~{\bf #1}, #3 (#2)}

\begin{document}
\removequeries

\begin{flushright}
LA-UR-06-4549 \\
RM3-TH/06-12 \\
June 2006
\end{flushright}

\vspace{0.8 true cm}
\begin{center}
{\Large {\bf  CP violation and Leptogenesis  \\ [3mm]
in  models with Minimal Lepton Flavour Violation}}\\
\vspace{1.0 true cm} {\large Vincenzo Cirigliano${}^{a,b}$,  Gino
Isidori${}^c$, Valentina Porretti${}^{d}$} \\
\vspace{0.5 true cm} ${}^a$
{\sl  Theoretical Division, Los Alamos National Laboratory, Los Alamos, NM 87545, USA} \\
${}^b$  {\sl  California Institute of Technology, Pasadena, CA 91125, USA}   \\
\vspace{0.1 true cm}
${}^c$ {\sl  INFN, Laboratori Nazionali di  Frascati,
   Via E. Fermi 40, I-00044 Frascati, Italy   } \\
\vspace{0.1 true cm}
${}^d$ {\sl    Dip. di Fisica, Univ. di Roma Tre,
   Via della Vasca Navale 84, I-00146 Roma, Italy } \\
\vspace{0.1 true cm}

\end{center}
\vspace{0.5cm}

\begin{abstract}
We investigate the viability of leptogenesis in models with
three heavy right-handed neutrinos, where
the charged-lepton and the neutrino Yukawa couplings
are the only irreducible sources of lepton-flavour
symmetry breaking (Minimal Lepton Flavour Violation hypothesis).
We show that in this framework a specific
type of resonant leptogenesis can be successfully accomplished. 
For natural values of the free parameters, this mechanism 
requires a high right-handed neutrino mass scale 
($M_\nu \gsim 10^{12}$ GeV). By means of a general effective field theory 
approach,  we analyse the impact of the CP violating
phases responsible for leptogenesis on the
low-energy FCNC observables and derive bounds on the scale of flavour violating new physics
interactions. As a result of the high value of the scale of total lepton-number violation,
in this class of models the $\mu\to e\gamma$ decay is expected to be close to the
present exclusion limit (under the additional assumption 
of new particles carrying lepton flavour at the TeV scale).
\end{abstract}

\section{Introduction}

All extensions of the Standard Model (SM)  with new degrees of
freedom at the TeV scale  carrying flavour quantum numbers have to
face the severe constraints implied by low energy Flavour Changing
Neutral Current (FCNC) transitions. An economical and elegant
solution to this {\em flavour problem} is provided by the Minimal
Flavour Violation (MFV) hypothesis, namely by the assumption that the
irreducible sources of  flavour symmetry breaking are minimally
linked to the fermion mass matrices observed at low energy. On the
one hand, the MFV hypothesis guarantees a suppression of FCNC rates
to a level consistent with experimental constraints without
resorting to unnaturally high scales of new physics.
On the other hand, this hypothesis provides a
predictive and falsifiable framework that links the possible
deviations from the SM in FCNC transitions to the measured fermion
spectrum and mixing angles.

The MFV  hypothesis  has a straightforward and unique realization in
the quark sector~\cite{Georgi,Hall:1990ac,MFV}: the SM Yukawa
couplings are the only sources of breaking of the $SU(3)_{Q_L}\times
SU(3)_{U_R}\times SU(3)_{D_R}$ quark-flavour symmetry.
The extension of the MFV hypothesis to the
lepton sector (Minimal Lepton Flavour Violation, MLFV) is less
straightforward: a proposal based on the assumption that the
breaking of total lepton number and lepton flavour are decoupled in
the underlying theory has recently been presented in
Ref.~\cite{Cirigliano:2005ck} and further analysed in
Ref.~\cite{Cirigliano:2006su}. The requirement
of minimality and predictivity has lead to the
identification of two independent MLFV scenarios \cite{Cirigliano:2005ck},
characterized by the different status
assigned to the effective Majorana mass matrix $m_\nu^{\rm eff}$
appearing as coefficient of the $|\Delta L| = 2$ dimension-five
operator in the low energy effective theory~\cite{Weinberg:1979sa}.

In the truly minimal case (dubbed {\em minimal field content}),
$m_\nu^{\rm eff}$, together with the charged-lepton Yukawa coupling $\lambda_e$, are assumed to
be the only irreducible sources of breaking of  $SU(3)_{L_L}\times SU(3)_{e_R}$
(the lepton-flavour symmetry of the low-energy theory).
One of the consequences of this hypothesis is the fact that
the only CP-violating phases involved in low-energy
FCNC observables are those contained in the lepton mixing matrix $\upmns$.
As a result, within this scenario it is not possible to address potential
correlations between low-energy observables and high-energy phenomena
such as leptogenesis~\cite{Fukugita:1986hr}. The viability of
leptogenesis crucially depends on the UV details of the model (see for
instance Refs.~\cite{rossi,D'Ambrosio:2004fz,Hambye:2005tk}),
which are beyond the control of the effective field theory approach.

In many realistic extensions of the SM, the flavour structure of the
theory is modified by the presence of  heavy right-handed neutrinos.
For this reason, a second scenario
(dubbed {\em extended field content}), with heavy
right-handed neutrinos and a larger lepton-flavour symmetry group,
$SU(3)_{L_L}\times SU(3)_{e_R} \times O(3)_{\nu_R}$, has also been
considered. In this extended scenario, the most natural and
economical choice about the symmetry-breaking terms is the
identification of the two Yukawa couplings, $\lambda_\nu$ and
$\lambda_e$, as the only irreducible symmetry-breaking structures,
in close analogy with the quark sector. In this context,
$m_\nu^{\rm eff} \sim  \lambda_\nu^T \lambda_\nu$ and the
lepton-number-breaking  mass term of the heavy right-handed
neutrinos is flavour-blind (up to Yukawa-induced corrections).

In the extended MLFV scenario the assumption about the
irreducible sources of lepton-flavour breaking does involve
the high energy sector of the theory. In particular, the symmetry
principle provides significant constraints on the amount of
CP violation in the decays of right-handed neutrinos, and thus
on leptogenesis. This could allow to establish some links between
low-energy FCNC observables and leptogenesis.
The investigation of these links is the main purpose of this work.
We address in particular the following questions:

\begin{itemize}

\item
Is leptogenesis viable  in models
where  the only sources of flavour breaking are  proportional to the
charged-lepton and neutrino Yukawa matrices  $\lambda_e$ and
$\lambda_\nu$? In other words: can one build nontrivial CP violating
re-phasing invariants using only  $\lambda_{\nu,e}$, and do they
contribute to the  CP asymmetries relevant for leptogenesis? This
question has implications
beyond the MLFV framework and
our findings provide a significant extension to the existing
statements in the literature~\cite{Hambye:2004jf}.

\item
In presence of CP violation, what is the flavour structure of the effective FCNC
couplings? Lifting the technical assumption of CP conservation which
was previously invoked to gain predictive
power~\cite{Cirigliano:2005ck,Cirigliano:2006su},
is the framework still predictive?

\item
Does the requirement of successful leptogenesis reduce
the large uncertainty on the scale of lepton-number
violation? Does this help to reduce the overall uncertainty
in the predictions of low-energy FCNC rates?
\end{itemize}

In Section~\ref{sect:CPV}, after introducing the basic structure of the model,
we show that the first of the above questions has a positive answer.
We indeed find nontrivial
re-phasing invariants that are in one-to-one correspondence with
the CP asymmetries in the decays of the quasi-degenerate
heavy Majorana neutrinos. Therefore, leptogenesis is in principle
viable within this framework.

Next  we  perform a numerical  study of leptogenesis
(Sect.~\ref{sect:analysis}) and analyse the implications for
charged-lepton FCNC processes (Sect.~\ref{sect:FCNC}). We find that
leptogenesis is also phenomenologically acceptable
within this framework. 
In general, the presence of new CP-violating
(CPV) phases leads to non-trivial modifications of the pattern of
FCNC rates obtained in the CP conserving limit. 
However, we also find that there is an interesting regime of small CPV phases
where the FCNC pattern is dictated again by the neutrino
oscillation parameters. 
Within this regime, the flavour structure
of the effective FCNC couplings receives small corrections
with respect to the CP conserving case and the
framework is particularly predictive.
  
The most interesting outcome of the numerical
study of leptogenesis in the MLFV framework is an approximate 
lower bound on  the right-handed neutrino mass: $M_\nu \gsim 10^{12}$ GeV, 
for natural values of the free parameters.  This allows us
to address the third question: we find
that the requirement of both successful leptogenesis
and FCNC rates in agreement with experiments
leads to non-trivial constraints on the overall scales of new physics
(more precisely the scales of lepton-number and lepton-flavour breaking).
In particular, we strengthen the conclusion of
Ref.~\cite{Cirigliano:2005ck} that within MLFV models
the $\mu\to e\gamma$ decay is expected to be within the
reach of the MEG experiment~\cite{MEG}.

\section{CP violation in models with MLFV and heavy $\nu_R$}
\label{sect:CPV}

\subsection{Framework}
\label{sect:frame}

The  {\em extended field content}  MLFV scenario postulates the
existence --beyond the Standard Model (SM) degrees of freedom-- of
three right-handed neutrino fields  singlets under the  SM gauge
group. The right-handed neutrino mass  (the only source of
$U(1)_{\rm LN}$  breaking) is flavour-blind  and its  scale is large
compared to the electroweak symmetry breaking scale ($M_\nu \gg v$):
\beq
{\cal L}_{(0)}  \supset  \ {\cal L}_{\rm gauge}^{\rm SM} +
 \frac{1}{2}~(M^{(0)}_R)_{ij}~
\bar \nu^{ci}_R\nu_R^j+\hc\qquad\text{with}\qquad (M^{(0)}_R)_{ij}
=M_\nu \,  \delta_{ij}~.
\label{eq:one}
\eeq
The lepton flavour symmetry  group of ${\cal L}_{(0)}$ is $G_{\rm
LF}= SU(3)_{L_L}\times SU(3)_{e_R}  \times O(3)_{\nu_R}$ and the
MLFV hypothesis states  that $G_{\rm LF}$ is broken only by two
irreducible sources, $\lambda_e^{ij}$ and $\lambda_\nu^{ij}$,
defined by:
\begin{align}
\label{eq:extlag} \CL_{\text{Sym.Br.}}  & \supset - \lambda_e^{ij}
\,\bar e^i_R(H^\dagger L^j_L) +i \lambda_\nu^{ij}\bar\nu_R^i(H^T
\tau_2L^j_L)+\hc
\end{align}
where $H$ is the usual SM Higgs doublet. Treating $\lambda_{e,\nu}$
as spurions of $G_{\rm LF}$, this implies the following
transformation properties:
\bea
& \lambda_e \to
V_R^{\phantom{\dagger}} \,\lambda_e V_L^\dagger~, \qquad \lambda_\nu
\to O_\nu \,\lambda_\nu V_L^\dagger~,& \label{eq:sp2}
\eea
with $V_L\in SU(3)_{L_L}$, $V_R\in SU(3)_{e_R}$, and $O_\nu\in
O(3)_{\nu_R}$. All the operators of the effective theory should
respect a formal invariance under $G_{\rm LF}$ according to these
rules.

Although in the effective field theory approach we remain agnostic
as to the full structure of ${\cal L}_{(0)}$ and
$\CL_{\text{Sym.Br.}}$ (as well as  the mechanism generating
$M_\nu$, $\lambda_e$,  $\lambda_\nu$), the MLFV postulates,
implemented through the spurion technique, provide enough
information to address specific questions concerning:

\begin{enumerate}

\item
{\em The structure of low energy FCNC couplings.}\\
The effective coupling governing FCNC transitions of charged leptons
to leading order in $\lambda_e, \lambda_\nu$, --or the leading
$(8,1,1)$ spurion of $G_{\rm LF}$-- is~\cite{Cirigliano:2005ck,Cirigliano:2006su}
\beq \Delta_{\rm FCNC}
= \lambda_\nu^\dagger \lambda_\nu~.
\eeq

\item
{\em The structure of $\nu_R$  mass splitting.}\\
The mass degeneracy of the $\nu_R$ fields  is removed by appropriate
combinations of spurions transforming as $(1,1,6)$ under $G_{\rm
LF}$. To lowest order in $\lambda_e$ and $\lambda_\nu$, we can write
\beq
M_R = M_R^{(0)} + \sum c_{n}~ \delta M_R^{(n)}~,
\label{eq:MR_eff}
\eeq
where $M_R^{(0)}$ has been defined in Eq.~(\ref{eq:one}) and
\begin{eqnarray}
\delta M_R ^{(11)}  &=& M_\nu  \left[   \lambda_\nu
\lambda_\nu^\dagger +  (\lambda_\nu \lambda_\nu^\dagger)^T  \right]
\ ,
\nonumber \\
\delta M_R ^{(21)}  & = &  M_\nu
 \left[  \lambda_\nu \lambda_\nu^\dagger  \lambda_\nu \lambda_\nu^\dagger
+  (\lambda_\nu \lambda_\nu^\dagger  \lambda_\nu \lambda_\nu^\dagger
)^T  \right]  \ ,
\nonumber \\
\delta M_R ^{(22)}  & = &  M_\nu  \left[   \lambda_\nu
\lambda_\nu^\dagger (\lambda_\nu \lambda_\nu^\dagger)^T  \right]  \
,
\nonumber \\
\delta M_R ^{(23)}  & = &  M_\nu  \left[  (\lambda_\nu
\lambda_\nu^\dagger)^T \lambda_\nu \lambda_\nu^\dagger   \right]  \
,
\nonumber \\
\delta M_R ^{(24)} & = &  M_\nu  \left[  \lambda_\nu
\lambda_e^\dagger  \lambda_e \lambda_\nu^\dagger +  (\lambda_\nu
\lambda_e^\dagger  \lambda_e  \lambda_\nu^\dagger )^T  \right]   \ ,
\nonumber \\
\delta M_R ^{(31)} &= & .... \label{eq:splitting1}
\end{eqnarray}
The $c_n$ are arbitrary coefficients whose size depends on
dynamical properties: if the Yukawa corrections are generated
within a perturbative regime, such as in scenarios of radiative
leptogenesis \cite{GonzalezFelipe:2003fi}, the size of the $c_{n}$ 
decreases according to the power of Yukawa insertions
(e.g.~in a standard loop-expansion one expects  $c_{11} \sim g_{\rm eff}^2/(4 \pi)^2$,
$c_{2i} \sim c^2_{11}$, \ldots). A priori one cannot exclude
a strong-interaction regime where all the  $c_{n} \sim \cO(1)$.
But even in this extreme case, the series in (\ref{eq:MR_eff}) is
expected to be dominated by the first few terms if the largest
entries of $\lambda_{\nu,e}$ are at most of $\cO(1)$
(as assumed in Ref.~\cite{Cirigliano:2005ck}
and expected in most scenarios).

\item  {\em The structure of CP asymmetries in $\nu_R$ decays relevant to leptogenesis.} \\
Denoting by $N_{1,2,3}$  the heavy neutrino mass eigenstates,
with masses $M_{1,2,3}$ determined by diagonalization
of $M_R$, the CP asymmetries $\epsilon_i$ relevant to leptogenesis
are \cite{CPasymmetries,Ellis,Lepto2}
\beq
\epsilon_i  \equiv \frac{ \sum_k  \left[  \Gamma (N_i \to l_k
H^*) - \Gamma (N_i \to \bar{l}_k  H) \right] }{ \sum_k \left[ \Gamma
(N_i \to l_k H^*) + \Gamma (N_i \to \bar{l}_k  H) \right] } =
-\sum_{j\neq
i}\frac{3}{2}\frac{M_i}{M_j}\frac{\Gamma_j}{M_j}I_j\frac{2 S_j +
V_j}{3} ~,
\label{eq:epsi}
\eeq
where
\beq
S_j=\frac{M_j^2 \left(M_{j}^2-M_{i}^2\right)
}{\left(M_{j}^2-M_{i}^2\right)^2 + M_i^2\Gamma_J^2} \qquad
V_j=2\frac{M_j^2}{M_i^2}\left[\left(1+\frac{M_j^2}{M_i^2}\right){\rm
log}\left( 1+\frac{M_i^2}{M_j^2}\right)-1\right]\label{self}
\eeq
and
\beq
I_j=\frac{{\rm Im}\left[\left(\bar{\lambda}_\nu
\bar{\lambda}_\nu^\dagger\right)^2_{ij}\right]}{|\bar{\lambda}_\nu\bar{\lambda}_\nu^\dagger|_{ii}|
\bar{\lambda}_\nu\bar{\lambda}_\nu^\dagger|_{jj}} \qquad\frac{\Gamma_j}{M_j}=\frac{|
\bar{\lambda}_\nu \bar{\lambda}_\nu^\dagger|_{jj}} {8\pi}~.
\label{eq:Ij}
\eeq
The factors $S_{j}$ and $V_{j}$ arise
respectively from one-loop self-energy and vertex
contribution to the decay widths of the heavy neutrinos
(for quasi degenerate right-handed neutrinos, $S\gg V$).

In Eqs.~(\ref{eq:epsi})--(\ref{eq:Ij})
$\bar{\lambda}_{\nu}$ indicates the neutrino Yukawa
coupling in the basis where $M_R$ is diagonal. 
Denoting
by $\bar U$ the unitary matrix
that diagonalizes $M_R$, one has
$\bar{\lambda}_{\nu} = \bar U  \lambda_{\nu}$. 
It is clear that
the size of the $\epsilon_i$ is determined by the misalignment
between $M_R$ and the spurion
\bea
h_\nu\,\equiv\,\lambda_\nu \lambda_\nu^\dagger\;.
\label{eq:hnu}
\eea
The key issue at this point is whether or not this
misalignment can be generated if the only sources
of flavour breaking are proportional to $\lambda_\nu$
and  $\lambda_e$. We show below that it is actually possible,
even in the limit $\lambda_e \to 0$.
\end{enumerate}

\subsection{Counting and characterizing CP violating phases}
\label{sec:invar}

The independent CP-violating phases of the model can be characterized in terms
of weak-basis invariants, i.e.~quantities that are insensitive to changes of
basis or re-phasing of the lepton fields.
Using the technique of Ref.~\cite{Santamaria:1993ah} or
Refs.~\cite{Bernabeu:1986fc,Branco:1989bn,Branco:2001pq}, one can
easily check that MLFV  models contain  six  independent
CPV invariants coming from the Yukawa sector.\footnote{~In a
specific underlying model there might be additional
flavour-conserving CPV phases --related to couplings of
heavy degrees of freedom-- on which our effective field theory
approach remains agnostic.} Even in the more restrictive case of
$\lambda_e = 0$  the model contains three independent CPV invariants.

In order to identify the weak-basis invariants,
one defines the most general CP
transformation as follows~\cite{Bernabeu:1986fc}:
\beq
\begin{array}{lcrclcr}
 \nu_L   &  \rightarrow &  U \,   C \nu_L^*~,  & \qquad\qquad
&  e_L   &  \rightarrow & U \,   C  e_L^*~,   \\
   e_R   &  \rightarrow &  V \,   C  e_R^*~,   & \qquad\qquad
& \nu_R   &  \rightarrow &  W \,   C  \nu_R^*~,
\end{array}
\eeq
where $U,V,W$ are unitary matrices. The above definition
corresponds  to a combination of CP and the most general flavour
transformation that leaves invariant the gauge-kinetic term.
%
%
In presence of a generic Majorana mass term $M_R$ for $\nu_R$  and
Yukawa interactions $\lambda_{e,\nu}$,  the  theory  is CP invariant
if and only if  there exists a set of $U,V,W$ such that:
\begin{eqnarray}
V^\dagger   \, \lambda_e \, U & =&  \lambda_e^*~, \\
W^\dagger   \, \lambda_\nu \, U & =&  \lambda_\nu^*~, \\
W^T   \, M_R  \, W  & =&  -M_R^*~.
\end{eqnarray}
%
%
%
%
Given this,  the simplest necessary conditions for CP invariance can
be cast in the following  weak-basis invariant
form~\cite{Branco:2001pq}
\begin{eqnarray}
B_1 & \equiv & {\rm Im} \,
{\rm Tr} \ \left[   h_\nu  \, (M_R^\dagger M_R)   M_R^*   \, h_\nu^*  M_R  \right] = 0~,  \\
B_2 & \equiv & {\rm Im} \,
{\rm Tr} \ \left[   h_\nu  \, (M_R^\dagger M_R)^2   M_R^*   \, h_\nu^*  M_R  \right] = 0~,  \\
B_3 & \equiv & {\rm Im} \, {\rm Tr} \ \left[   h_\nu  \,
(M_R^\dagger M_R)^2   M_R^*   \, h_\nu^*  M_R (M_R^\dagger M_R)
\right] = 0  \ , \label{eq:invariants}
\end{eqnarray}
where we used the definition of $h_\nu$ in Eq.~(\ref{eq:hnu})
and $M_R$ denotes a generic heavy neutrino mass term.
The invariants $B_{1,2,3}$ are independent and one can
construct  three other independent invariants that  explicitly
involve $\lambda_e$, by replacing  $h_\nu  \to h_e \equiv
\lambda_\nu \lambda_e^\dagger \lambda_e \lambda_\nu^\dagger $ in the
expressions of $B_{1,2,3}$.
Moreover, $B_{1,2,3}$ are in direct correspondence with the CP
asymmetries $\epsilon_i$ relevant for leptogenesis, as can be seen
by expressing them in the weak-basis where $M_R$ is diagonal with
eigenvalues $M_{1,2,3}$. In this basis, for example, $B_1$ reads:
\begin{eqnarray}
B_1 &=&  M_1 M_2 (M_2^2 - M_1^2)  \ {\rm Im} \left( (\bar
\lambda_\nu \bar \lambda_\nu^\dagger)_{12}^2 \right) +
  M_1 M_3 (M_3^2 - M_1^2)
 \  {\rm Im} \left( (\bar \lambda_\nu \bar \lambda_\nu^\dagger)_{13}^2 \right)
   \nonumber \\
  &+&    M_2 M_3 (M_3^2 - M_2^2)
 \  {\rm Im} \left( (\bar \lambda_\nu \bar \lambda_\nu^\dagger)_{23}^2 \right)~.
\end{eqnarray}

Let us now investigate whether the $B_{i}$ are not vanishing
with specific structures of $M_R$ satisfying the MLFV hypothesis:

\begin{itemize}
\item If $M_R $ is proportional to the identity, the hermiticity of $h_\nu$ and the
cyclic property of the trace operation imply that the $B_{i}$
vanish identically.

\item  The next step is to break the degeneracy of the heavy neutrinos in a way
consistent with the MLFV hypothesis, using the $\delta M_R^{(i)}$ in
Eq.~(\ref{eq:splitting1}). If we stop to terms quadratic in  the
Yukawa couplings, namely $M_R =   M_\nu  I  +  c_{11}  \, \delta
M_R^{(11)}$, the $B_{i}$ are still vanishing (again because of the
hermiticity of $h_\nu$ and the trace properties). This is consistent
with the findings of Ref.~\cite{Hambye:2004jf}, where only the
mass-splitting $\delta M_R^{(11)}$ was considered.

\item  The vanishing of $B_{1,2,3}$ is finally avoided
if any of the quartic terms in  Eq.~(\ref{eq:splitting1}) is
considered. This can be verified in a number of ways. The simplest
method is to use an explicit parameterization of $\lambda_e$ and
$\lambda_\nu$  (as  the one given in the next subsection) to evaluate
the traces in the expression of $B_{1,2,3}$.

\item
It is worth stressing that $B_{1,2,3} \neq 0$ even in the limit
$\lambda_e \to 0$. We will show in Sect.~\ref{sect:analysis} that
 successful leptogenesis can actually be
accomplished in this regime.
For instance, the structure $M_R =   M_\nu  I  +
c_{21} ~ \delta M_R^{(21)}$  is sufficient to generate non-vanishing
CP invariants. We conclude that the vanishing of
$B_{1,2,3}$ up to $\cO(\lambda_\nu^2)$ is just an accidental
cancellation. This fact can be understood by noting that in absence
of $\lambda_e$ the simplest non-vanishing invariants have a structure
of the type
\beq B_0\,=\,{\rm Im} \, {\rm Tr} \ \left[   (h_\nu)^a
(h_\nu^b)^*  (h_\nu)^c (h_\nu^d)^*  \right]~,
\label{eq:B00}
\eeq
with $(a,b,c,d)$ non zero and $b\not=d$,\,$a\not=c$.
This structure appears in the
$B_{1,2,3}$ only if $M_R$ is at least quartic in $\lambda_\nu$.
\end{itemize}

In conclusion our general analysis of weak-basis
invariants\footnote{~Although in our discussion  we have used as
starting point the well established  invariants introduced in
Ref.~\cite{Branco:2001pq}, it is possible to  reformulate the above
argument entirely in terms of $\lambda_\nu$ and $\lambda_e$, without
any reference to the mass matrix $M_R$ of right-handed neutrinos.}
shows that leptogenesis is {\it at least in principle} possible
within MLFV. In the next section we perform a  quantitative study of
the asymmetries in order to identify the regions of parameter space in
which the baryon asymmetry generated via leptogenesis matches the
observed value.

\subsection{Explicit parametrization of $\lambda_\nu$}
\label{sect:param}

In order to investigate the phenomenological consequences of the
MLFV framework in presence of CP violation, it is convenient to
choose a particular weak basis and a parameterization of
$\lambda_e$, $\lambda_\nu$.

In the basis where the charged-lepton mass matrix is diagonal, we
can write
\bea
m_\nu^{\rm eff} \equiv  v^2 \lambda_\nu^T M_R^{-1}\lambda_\nu =
\ustapmns m_{\rm diag} \udagpmns \label{eq:mDelta}
\eea
where $\upmns$ is the mixing matrix of the light neutrinos
and the corresponding mass eigenvalues are encoded in
$m_{\rm diag }={\rm diag}(m_{\nu_1},m_{\nu_2},m_{\nu_3})$.
In this basis, the most general form of $\lambda_\nu$ is \cite{Casas:2001sr}
\beq
\lambda_\nu = \frac{1}{v} M_R^{1/2} R~ m_{\rm
diag}^{1/2} ~\udagpmns  \quad \stackrel{\rm MLFV}{\longrightarrow} \quad
\frac{M_\nu^{1/2}}{v}\,R~ m_{\rm diag}^{1/2} ~\udagpmns
 \label{eq:R}
\eeq
where $R$ is a generic complex orthogonal matrix: $R^T R =1$.
As explicitly indicated in the last expression of Eq.~(\ref{eq:R}),
within the MLFV framework $M_R$ is to a very good approximation
proportional to the identity matrix ($M_R \approx M_\nu \times I$).
The small Yukawa-induced mass-splittings, which
are crucial for leptogenesis, can be safely neglected in the
reconstruction of the see-saw relation.

The matrix $R$ contains six real parameters and can be decomposed as
\beq R= O \times H \eeq where $O$ is a real orthogonal matrix and
$H$ a complex orthogonal and hermitian matrix
\beq\label{eq:Hprop}
H^T H =1 ~, \qquad H^\dagger = H~.
\eeq
 Both $O$ and $H$ can be
expressed in terms of 3 independent parameters. Thanks to the
$O(3)_{\nu_R}$ invariance --independently of any assumption about CP
invariance-- within the MLFV framework we  can always choose a basis of right-handed fields
such that $O=1$. Our irreducible parameterization of $\lambda_\nu$
is then
\beq
\lambda_\nu = \frac{M^{1/2}_\nu}{v}~H~m_{\rm diag}^{1/2}~\udagpmns~.
\label{eq:lpar}
\eeq
 In the CP limit, $H=I$. The CPV nature of H is transparent in the following
parameterization~\cite{Pascoli:2003rq},
\beq
\label{eq:phi}
H = e^{i
\Phi}= I - \frac{\cosh r - 1}{r^2} \, \Phi^2  + i \frac{\sinh r}{r}
\, \Phi~ , \qquad \Phi= \left(\begin{array}{ccc}
0 & \phi_1 & \phi_2 \\
- \phi_1 & 0 &  \phi_3 \\
- \phi_2 & - \phi_3 & 0
\end{array} \right)~,
\eeq
where the $\phi_i$ are real parameters and
$r=\sqrt{\phi_1^2+\phi_2^2+\phi_3^2}$. The CP-conserving case
analysed in Ref.~\cite{Cirigliano:2005ck,Cirigliano:2006su} is
recovered in the limit\footnote{~In the following 
we will often refer to the $\phi_i$ as to the CPV phases of $H$;
this notation is a bit misleading since the $\phi_i$
also control the modulus of $H$.} $\phi_i\to 0$.

Using the parameterization (\ref{eq:lpar}), the strength of FCNC
couplings reads
\beq
\Delta_{\rm FCNC} = \lambda_\nu^\dagger \lambda_\nu =
\frac{M_\nu}{v^2}~\upmns ~ m_{\rm diag}^{1/2}~ H^2~
m_{\rm diag}^{1/2}~\udagpmns \label{eq:fcnc}
\eeq
and  depends not only on the neutrino mass spectrum and mixing
angles, but also on the CPV phases appearing in $\upmns$ and in $H$.
On the other hand, $H$ also controls the CPV phases linked to leptogenesis:
\beq
 h_\nu\,\equiv \,\lambda_\nu \lambda_\nu^\dagger = \frac{M_\nu}{v^2}~H~ m_{\rm diag}
  ~H \label{eq:lept}  ~.
\eeq
Eqs.~(\ref{eq:fcnc}) and (\ref{eq:lept})  reveal an interesting connection
between the CPV phases relevant to leptogenesis  and the effective
strength of LFV processes that will be explored in Section~\ref{sect:FCNC}.
Note that even for small $\phi_i$ the results for $\Delta_{\rm FCNC}$
obtained in the CP-conserving case can be a poor approximation, 
since the hyperbolic functions in $H$ rapidly become large, 
and even small off-diagonal entries in $H$ can spoil the 
hierarchical structure of $m_{\rm diag}$.

Before concluding this section, let us comment on how the
CPV invariants $B_i$ discussed in Section~\ref{sec:invar} depend
on the CPV phases of $H$.
Expanding to the first non-trivial order in the CPV
parameters $\phi_i$, the basic invariant $B_0$,
--computed for $(a,b,c,d)=(1,2,2,1)$ according to Eq.~(\ref{eq:B00})--
reads
\beq
B_0 \, \propto \, \phi_1\phi_2\phi_3 \left(m^2_{\nu_1}-m^2_{\nu_2}\right)
\left(m^2_{\nu_1}-m^2_{\nu_3}\right) \left(m^2_{\nu_2}-m^2_{\nu_3}\right)~.
\eeq
As we will discuss in Section~\ref{sect:analysis}, this result has
phenomenological interesting consequences: it implies that in
limit $\lambda_e\rightarrow 0$, all the three phases of $H$
must be non vanishing in order to generate a sufficient
amount of CP violation.

\section{Numerical analysis of the leptogenesis conditions}
\label{sect:analysis}

Following the standard leptogenesis
analyses~\cite{Lepto2},
we assume that the observed baryon asymmetry of the Universe, here
defined in terms of the baryon, antibaryon and photon number
densities,
\beq \eta_B\, =\,\frac{n_B - n_{\bar B}}{n_\gamma}\,=\,
(6.3\pm 0.3)\times 10^{-10}~,
\label{eq:YBexp}
\eeq
entirely originates from a lepton
asymmetry, $\eta_L$, transferred to the baryon sector via sphaleron
processes \cite{Fukugita:1986hr}.

The matter-antimatter asymmetries $\eta_L$ and $\eta_B$ are connected by
an $\cO(1)$ factor which depends on the spectrum of the theory.
Working in a generic effective field theory approach, we are not
able to determine this factor, as well as many other specific
details of the model. Our goal is the determination
of the main conditions necessary to achieve the correct order
of magnitude of $\eta_B$. 
In particular, we assume
$\eta_B \approx - \eta_L/2$
as in the SM and consider the range [ $3\times 10^{-10}\lesssim\eta_B\lesssim
9\times 10^{-10}$ ] as phenomenologically acceptable.
 The baryon asymmetry $\eta_B$ can be expressed as
\beq
\eta_B = 0.0096 \ \sum_{i} \, \epsilon_{i} \, d_{i}
\eeq
where $d_i$ are the washout factors, and the $\epsilon_i$ are the CP
asymmetries defined in Eq.~(\ref{eq:epsi}). The washout functions
$d_i$ can be found solving numerically the full set of Boltzmann
equations. According to the recent analysis of Ref.~\cite{DiBari},
in the case of three quasi degenerate right-handed neutrinos
the washout factors are approximately equal and can be expressed as
\beq
d_i = d (K_1 + K_2 + K_3)  \qquad
d(K)=\frac{2}{K z_B(K)}\left(1-e^{-\frac{K
z_B(K)}{2}}\right)
\label{eq:di}
\eeq
where the $K_i$ are the decay parameters
defined as the total decay widths of the $\nu_R^i$  normalized to
the expansion rate at $T=M_\nu$ and\footnote{~The results 
in Eqs.~(\ref{eq:di})--(\ref{eq:di2}) provide an excellent 
representation of the fully numerical solutions in the strong washout 
regime, in which $K_i > 1$ (i=1,2,3).
We have checked that in our framework this condition
is verified in a large fraction of the parameter space.}
\beq
z_B(K) \approx 2+4
K^{0.13}e^{-\frac{2.5}{K}}\;.
\label{eq:di2}
\eeq

\begin{figure}[t]
\begin{center}
\includegraphics[angle=270,width=7cm]{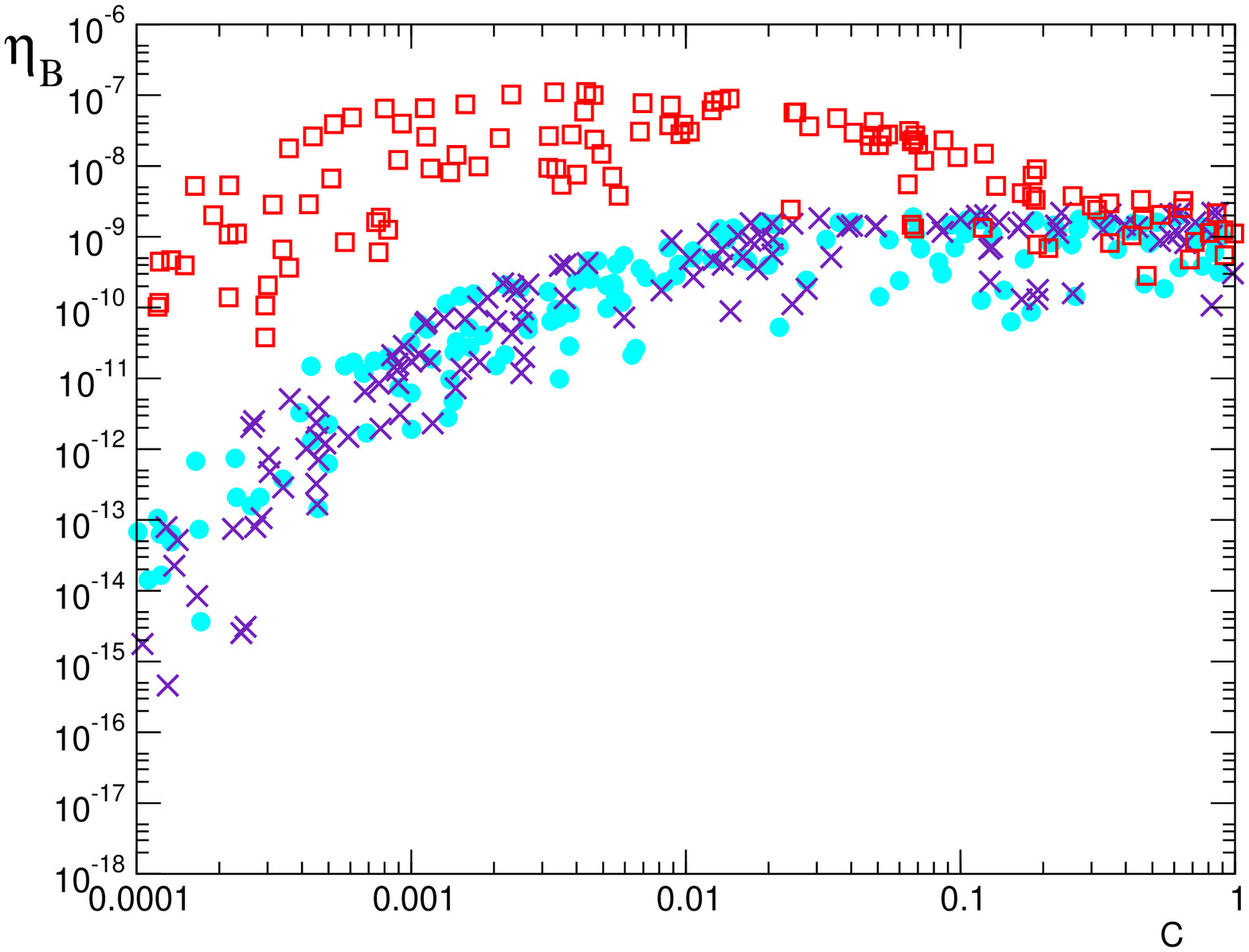}
\hspace{0.5 cm}
\includegraphics[angle=270,width=7cm]{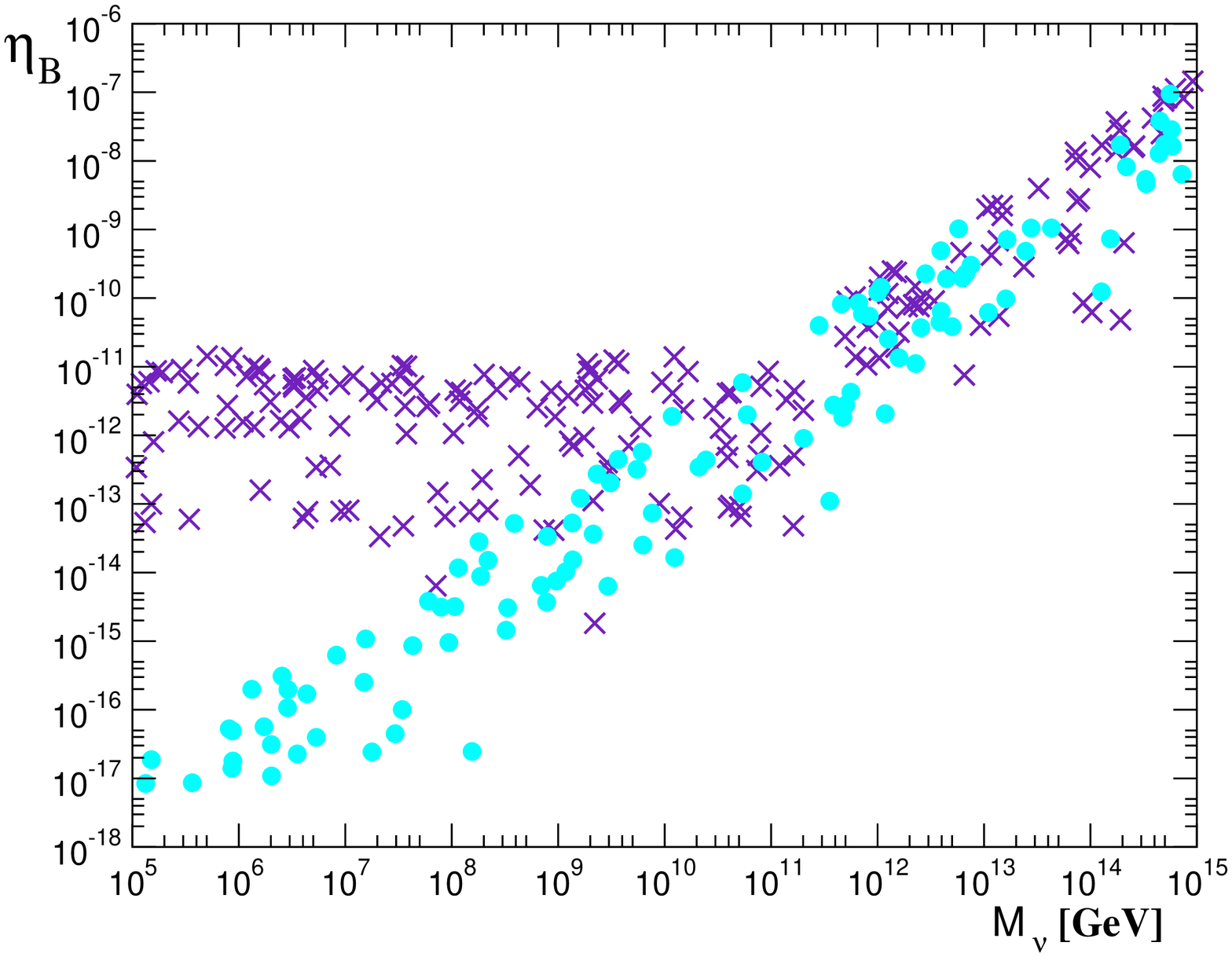}
\caption{\label{fig:one} Dependence of $\eta_B$ on the size of the
mass-splitting parameter ($c$) and on the absolute mass scale ($M_\nu$) of
the $\nu_R$ mass matrix. {\em Left panel}: $\eta_B$ as a function of 
$c$. The light blue circles
correspond to the reference hierarchy in Eq.(\ref{cases}); the violet
crosses are obtained for $c^{(24)} = 0$; the red squares are
obtained for  $c^{(11)} = c^{(21)} = c^{(24)} = c$; in all cases
$M_\nu = 10^{13}$ GeV. {\em Right panel}: $\eta_B$ as a function of $M_\nu$.
Conventions as in the left
panel, with $c \in [0.001,0.1]$. In both plots $\phi_i \in
[0.01,0.6]$ and $m_\nu \in [10^{-4},10^{-2}]$~eV.}
\end{center}
\end{figure}

In order to explore the quantitative results of the framework built
in the previous section, we have to determine  the flavour structure of $M_R$
 specifying the coefficients
of the spurions which appear in Eq.~(\ref{eq:splitting1}). In most of our
numerical studies, we assume the following structure
\beq
\begin{array}{lclll}
& \qquad &  c^{(11)} = c & c^{(21)} = c^2 & c^{(24)} = c^2,  \\
\end{array}
\label{cases}
\eeq
with the other $c_n$ set to zero.
This assumption is quite general and presents
all the features relevant for our discussion.
For $c\ll 1$ we recover the natural expectation
of a perturbative regime, while a prototype of a
non-perturbative scenario is obtained
for $c \rightarrow 1$. We have explicitly checked
that the numerical results are essentially unaffected by
the substitution of $c^{(21)}$ with any $\cO(1)$ combination
of $\{ c^{(21)},c^{(22)},c^{(23)} \}$.\footnote{~There is only
one pathological choice, namely $c^{(21)}=c^{(22)}=c^{(23)}$.
In this case the quartic term in Eq.~(\ref{eq:splitting1})
is the square of the quadratic one and the CPV invariants
vanish identically as in the case with no quartic terms
(see Sect.~\ref{sec:invar}). } The most relevant features
of this assumption are illustrated in Fig.~\ref{fig:one},
where we show for comparison the limiting cases $c^{(24)}=0$
(i.e.~$\lambda_e\to 0$) and  $c^{(11)} = c^{(21)} = c^{(24)} = c$.

A key observation which holds in all cases is that the order of magnitude
of the mass splitting is dominated by $\delta M_R^{(11)}$, the first order 
term in  Eq.~(\ref{eq:splitting1}). This naturally yields to the resonance
condition~\cite{resonance} which enhances the self-energy
contribution:
\bea
(M_R^j - M_R^i)/M_R^i \,\sim\, |\bar\lambda_\nu\bar\lambda_\nu^\dagger|_{jj}\,\implies
 \,S_{j}\,\sim\,\frac{M_j}{\Gamma_j}\,\gg\,V_j~.
\eea
As a result, within the MLFV framework the general formula (\ref{eq:epsi})
can always be simplified to
\bea
\label{eq:simpl}
 \epsilon_i\,\sim\,-\,\sum_{j\neq i}\frac{{\rm Im}\left[\left(\bar{\lambda}_\nu
\bar{\lambda}_\nu^\dagger\right)^2_{ij}\right]}{|\bar{\lambda}_\nu\bar{\lambda}_\nu^\dagger|_{ii}|
\bar{\lambda}_\nu\bar{\lambda}_\nu^\dagger|_{jj}}\quad~.
\eea
The dependence of $\eta_B$ on the parameter $c$ which
determines the size of the mass-splittings
is illustrated in the left panel of Fig.~\ref{fig:one}.
As can be noted, the dependence is quite mild for $c\in[0.01,1]$
and employing the natural hierarchy (\ref{cases}).
The dependence is stronger for the case
$c^{(11)} = c^{(21)} = c^{(24)} = c$, which however is not
well motivated for $c \ll 1$.
(In Fig.~\ref{fig:one}, as in the following plots, the parameters which play a minor role
in illustrating a specific functional dependence have been fixed to
reference intervals specified in the captions).

The dependence of $\eta_B$ on $c_{24}$ --the coefficient of the
mass-splitting containing $\lambda_e$-- is well illustrated by
the right panel of Fig.~\ref{fig:one}:
the dependence is negligible if $M_\nu \gsim 10^{11}$~GeV
(see for instance the left panel for $M_\nu=10^{13}$~GeV),
while it is crucial for lighter values of $M_\nu$. This fact
can easily be understood by noting that the normalization
of $\lambda_\nu$ is proportional to $\sqrt  M_\nu$
[see Eq.~(\ref{eq:lpar})]. For sufficiently large values of
$M_\nu$ we enter the regime where the flavour-violating
asymmetry induced by the quartic terms in $\lambda_\nu$ is much
larger than the one induced by the mass splitting
containing $\lambda_e$. This also explains the growth of $\eta_B$
with $M_\nu$ for  $M_\nu \gsim 10^{11}$ GeV (see appendix).
Interestingly, the experimental 
value of $\eta_B$ in Eq.~(\ref{eq:YBexp})
indicates that the region of $M_\nu$
where the effects of $\lambda_e$ can be neglected is
the phenomenologically relevant one. In particular, 
varying the free parameters in what we consider to be
their natural range (see Fig.~\ref{fig:four}, upper panel), 
we find that a baryon asymmetry compatible with 
experiments implies  $M_\nu \gsim 10^{12}$ GeV.
Since we have not explicitly solved the full set of Boltzmann 
equations (we relied on the approximate formulae of Ref.~\cite{DiBari}),
and we have chosen a specific set of free parameters, 
this bound cannot be taken as a strict lower limit, 
but it should be interpreted as the natural lower scale 
for successful leptogenesis in this general framework. 
Note that, similarly to the negligible influence of $\lambda_e$ in the
initial values of the CPV asymmetries, in the high $M_\nu$ region
we can neglect other flavour effects which have been addressed
in the recent literature~\cite{flavour}.

\begin{figure}[t]
\begin{center}
\includegraphics[angle=270,width=7cm]{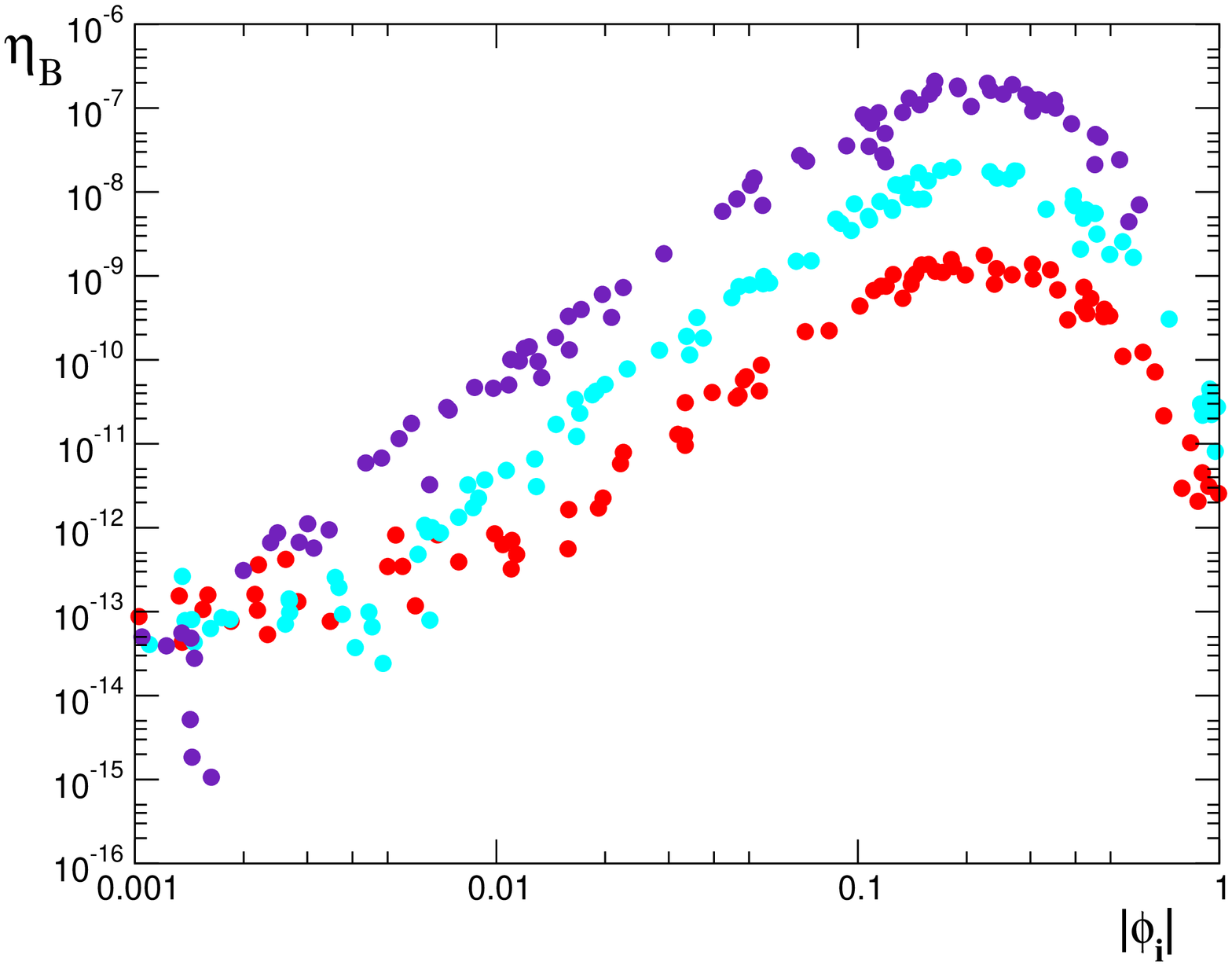}
\hspace{0.5 cm}
\includegraphics[angle=270,width=7cm]{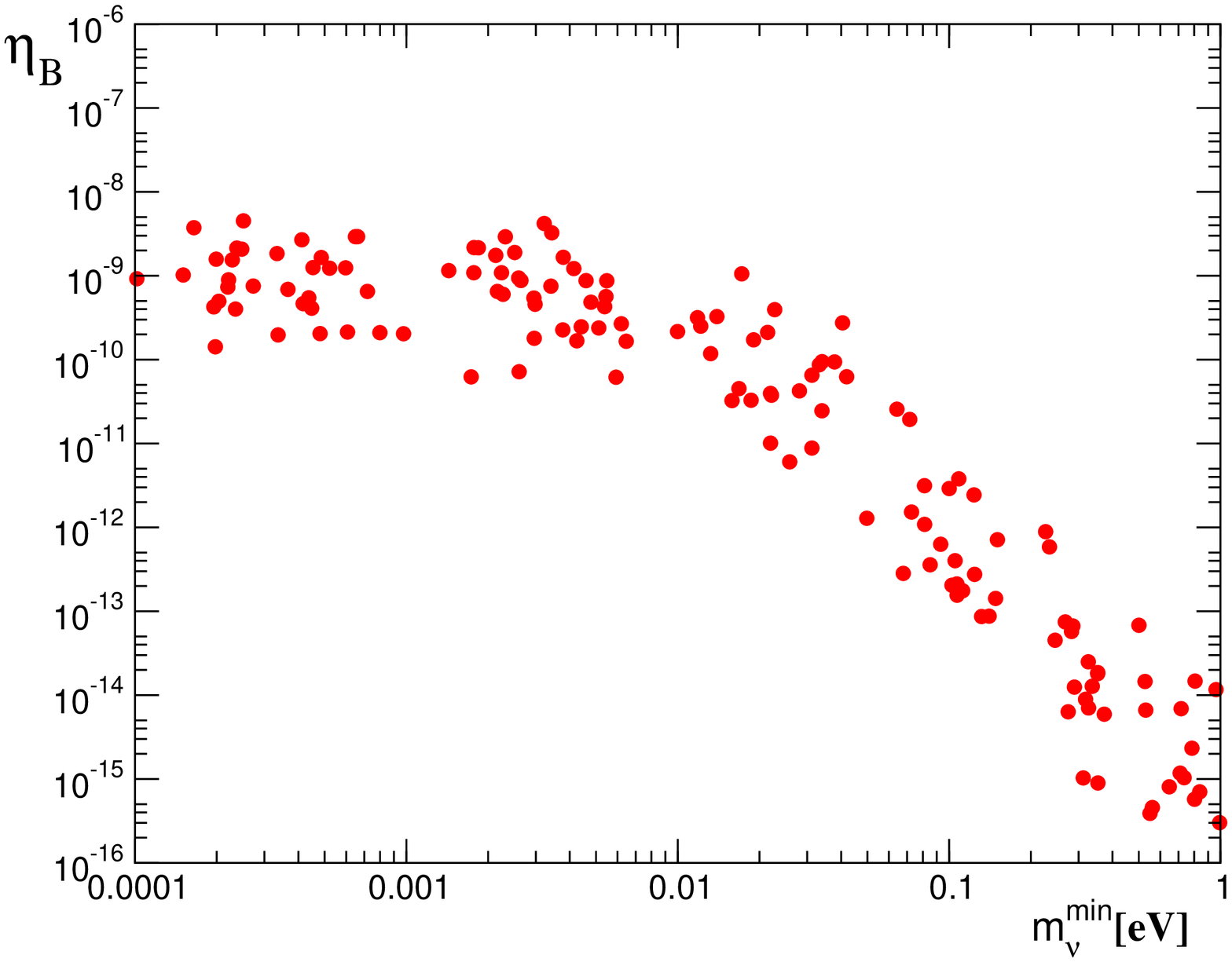}
\vspace{0.5 cm} \caption{\label{fig:two} Dependence of $\eta_B$ on the
CPV parameters of $\lambda_\nu$ ($\phi_i$) and on the mass of the lightest
left-handed neutrino ($m_\nu^{min}$). {\em Left panel}: $\eta_B$ as a function of
$\phi=\phi_1=\phi_2=\phi_3$ for $M_\nu=10^{13}$ GeV (red, lower
curve), $M_\nu=10^{14}$ GeV (light blue, middle), $M_\nu=10^{15}$
GeV (violet, upper), with $m^{min}_\nu=[10^{-4},10^{-2}]$~eV. {\em
Right panel}: dependence of $\eta_B$ on $m^{min}_\nu$, for
$M_\nu=10^{13}$ GeV and $\phi_i \in[0.1,0.6]$ (also the CP asymmetries $\epsilon_i$ 
show a similar dependence on $m_\nu^{min}$). In both plots
$c\in[0.01,0.1]$. Only the points satisfying $|\lambda_\nu|\lesssim
1$ are plotted.}
\end{center}
\end{figure}

The left panel of Fig.~\ref{fig:two}  illustrates the dependence of $\eta_B$
on the three CPV parameters $\phi_i$ contained in the matrix $H$.
In this plot the $\phi_i$  have been set to the same value. As mentioned
at the end of Sect.\ref{sect:param}, in the limit where
we can neglect $\lambda_e$, the CPV asymmetries depend on
the products of the three $\phi_i$. For this reason, the maximal
effects are obtained when the three  $\phi_i$ are similar in size.
The decrease of the baryon asymmetry for larger values
is due to a partial cancellation between the two terms in Eq.~(\ref{eq:simpl}),
as consequence of the peculiar properties of the matrix $H$ [see Eq.~(\ref{eq:Hprop})].
The suppression becomes more effective for large $\phi_i$ and for a
quasi-degenerate spectrum of the light neutrinos, as shown in the right panel of
Fig.~\ref{fig:two}. All plots are obtained assuming a normal hierarchy
for the light neutrinos, but we have checked that the results
are almost unchanged in the case of inverted hierarchy.

In conclusion, the numerical study shows that the correct order of
magnitude of the baryon asymmetry can be obtained in the
MLFV framework. The key ingredients are a sufficiently high overall
scale of right-handed neutrinos ($M_\nu \gsim 10^{12}$ GeV)
and sufficiently large CPV parameters in the matrix $H$
($|\phi_i| \gsim 0.01$).

\section{Implications for low-energy FCNC transitions}
\label{sect:FCNC}

In this section we briefly analyse the consequences on low-energy FCNC rates
following from the requirement of successful leptogenesis.
In particular, we are interested in understanding:
i) what are the implications on the overall rate of FCNC transitions;
ii) whether the predictions of ratios such as $\cB(\mu\rightarrow e\gamma)/\cB(\tau\rightarrow
\mu \gamma)$ derived in Ref.~\cite{Cirigliano:2005ck,Cirigliano:2006su}
in the limit of CP conservation are still valid.
 
We focus the attention on
$l_i\rightarrow l_j\gamma$ processes  only.
The ratio of the corresponding branching ratios depends uniquely
on the $\upmns$ matrix, the effective light neutrino masses and the
leptogenesis phases. Moreover, the absolute rates of these processes are 
particularly simple --compared to other FCNC transitions-- being determined 
by only two independent dimension-six effective operators.  
This allows us to assess in cleaner context the impact of the CPV parameters 
which have been neglected in Ref.~\cite{Cirigliano:2005ck,Cirigliano:2006su}.
Finally, significant experimental
improvements on both $\mu\rightarrow e\gamma$ and $\tau\rightarrow \mu(e)\gamma$
are expected in the near future \cite{FCNC_exp_rev}.
A detailed study of the dependence of different low-energy observables
on the leptogenesis and Majorana phases in a similar context
(SUSY see-saw with quasi-degenerate $\nu_R$) has been recently presented
in  Ref.~\cite{Petcov2}.

In the MLFV framework, the effective Lagrangian relevant for the
radiative decays $l_i\rightarrow l_j\gamma$ is
\bea
{\cal L}_{\rm eff}\,=\,\frac{1}{\Lambda_{\rm LFV}^2}\,\left(c^{(1)}_{RL}
O_{RL}^{(1)}\,+\,c^{(2)}_{RL} O_{RL}^{(2)}\right)~,
\eea
where
\bea
O_{RL}^{(1)}&\,=\,&g'H^\dagger\bar
e_R\sigma^{\mu\nu}\lambda_e\Delta_{\rm FCNC}L_LB_{\mu\nu}~, \no\\
O_{RL}^{(2)}&\,=\, &g H^\dagger\bar
e_R\sigma^{\mu\nu}\tau^a\lambda_e\Delta_{\rm FCNC}L_L W^a_{\mu\nu}~,
\eea
and $g'$ ($g$) and $B_{\mu\nu}$ ($W^a_{\mu\nu}$) are the coupling
constant and the field strength tensor of the $U(1)_Y$ ($SU(2)_L$)
gauge group. This effective Lagrangian leads to \cite{Cirigliano:2005ck}
\bea\label{eq:BR}
B_{l_i\rightarrow l_j\gamma} &\equiv&
\frac{ \Gamma (\ell_i \to \ell_j \gamma) }{
\Gamma (\ell_{i} \to \ell_{j} \nu_{i} \bar{\nu}_{j} )} ~=~
384\,\pi^2
e^2\,\frac{v^4}{\Lambda^4_{\rm LFV}}\,\left|\left(\Delta_{\rm FCNC\,}\right)_{ij}\right|^2\,
|c^{(2)}_{RL}-c^{(1)}_{RL}|^2  \no  \\
&=&
\left(\frac{v M_\nu }{\Lambda^2_{\rm LFV}}\right)^2~|c^{(2)}_{RL}-c^{(1)}_{RL}|^2 ~
\widehat f_{\ell_i \to \ell_j \gamma} \left( \upmns, m_{\nu}^{\rm min}, \phi_i \right)~.
\label{eq:Ri2}
\eea
The coefficients of the operators and the effective
new-physics scale $\Lambda_{\rm LFV}$ (expected to be
in the TeV region) are unknown, but they both
cancel in the ratios of the various FCNC rates.
For simplicity, we will set $|c^{(2)}_{RL}-c^{(1)}_{RL}|=1$
in the following.

\begin{figure}[p]
\begin{center}
\includegraphics[angle=270,width=10cm]{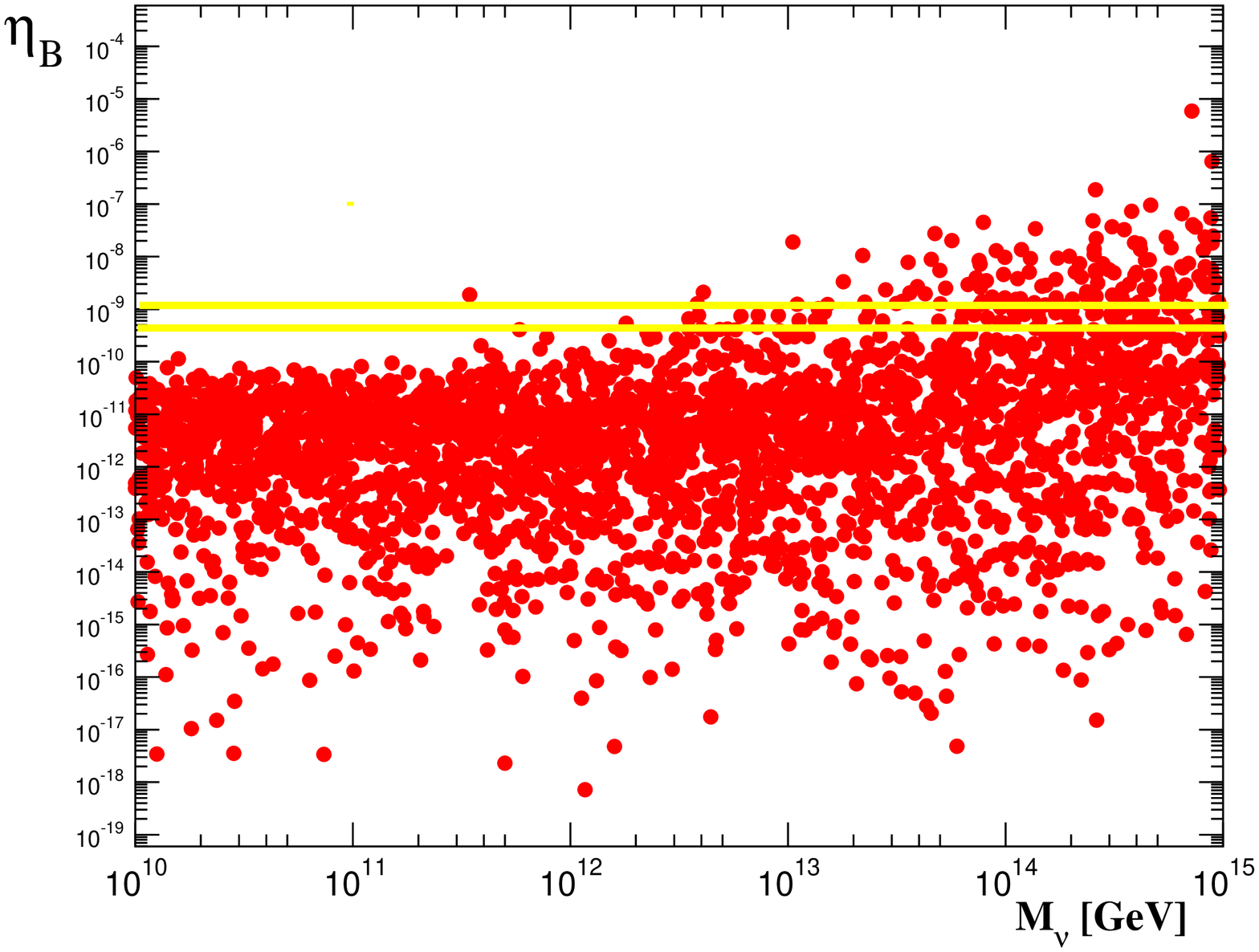}\\
\end{center}
\vspace{0.5 cm}
\begin{center}
\includegraphics[angle=270,width=10cm]{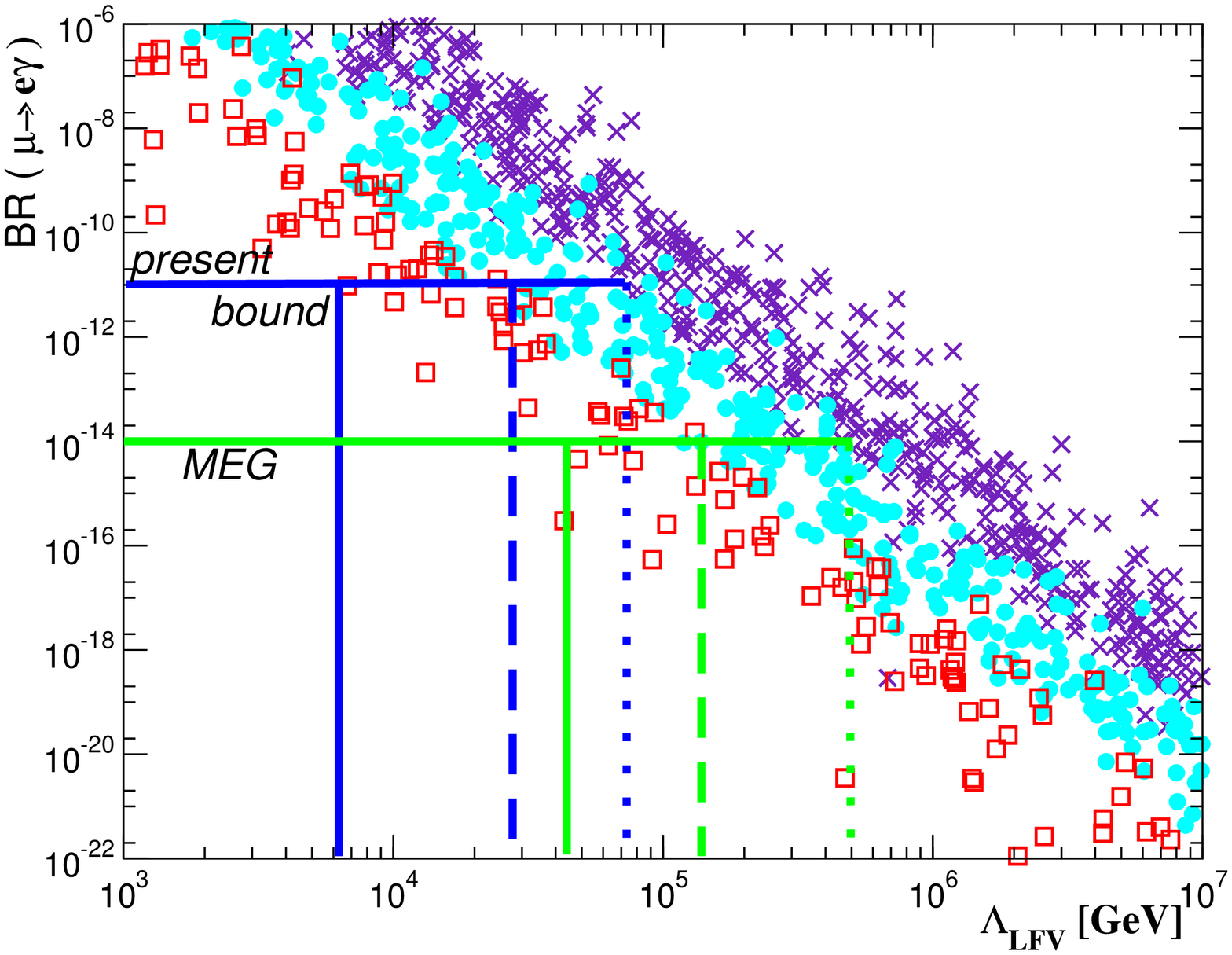}\\
  \caption{\label{fig:four} Numerical studies on the scales of lepton number
   ($M_\nu$) and lepton flavour ($\Lambda_{\rm LFV}$) violation.
    {\em Upper panel}: $\eta_B$ as a function of
$M_\nu$, varying the other parameters in the following ranges: $c
\in [0.001,1]$, $\phi_i \in [0.001,1]$ and $m_\nu^{min} \in
[10^{-4},0.6]$~eV. {\em Lower panel}: $\BR(\mu\to e\gamma)$ vs.
$\Lambda_{\rm LFV}$ for the points in the yellow band on the upper
panel (satisfying the leptogenesis constraint), grouped according to
their value of $M_\nu$: violet crosses for $M_\nu >10^{14}$ GeV,
light blue circles for $10^{14}$ GeV $>M_\nu >10^{13}$ GeV, red 
squares for $M_\nu <10^{13}$ GeV. The present  bounds (future MEG sensitivity) 
on $\Lambda_{\rm LFV}$ are shown by the blue (green) lines: full line
for $M_\nu=10^{13}$ GeV, dashed line for $M_\nu=10^{14}$ GeV, dotted
line for $M_\nu=10^{15}$ GeV. Without further information on $M_\nu$
and leptogenesis parameters, the bounds indicated by the full lines
should be taken as conservative estimates. }
\end{center}
\end{figure}

The most significant implication on FCNC rates derived from
the requirement of successful leptogenesis is the constraint on the
overall normalization of Eq.~(\ref{eq:Ri2}). As we have seen in the
previous section, the overall neutrino mass scale $M_\nu$ should
exceed $10^{12}$ GeV in order to generate the observed value of the
baryon asymmetry (see Fig.~\ref{fig:four}, upper panel). This breaks the
ambiguity in the normalization of $\lambda_\nu$ and
consequently on the ratio $(v M_\nu)/\Lambda^2_{\rm LFV}$
appearing in Eq.~(\ref{eq:Ri2}). In the analyses of
Ref.~\cite{Cirigliano:2005ck,Cirigliano:2006su}, the
ambiguity on the value of $M_\nu$ prevented the extraction of lower bounds
on the new-physics scale $\Lambda_{\rm LFV}$ using
the experimental constraints on FCNC rates. This becomes possible imposing
the additional requirement of successful leptogenesis.

\begin{figure}[p]
\begin{center}
\includegraphics[angle=270,width=7cm]{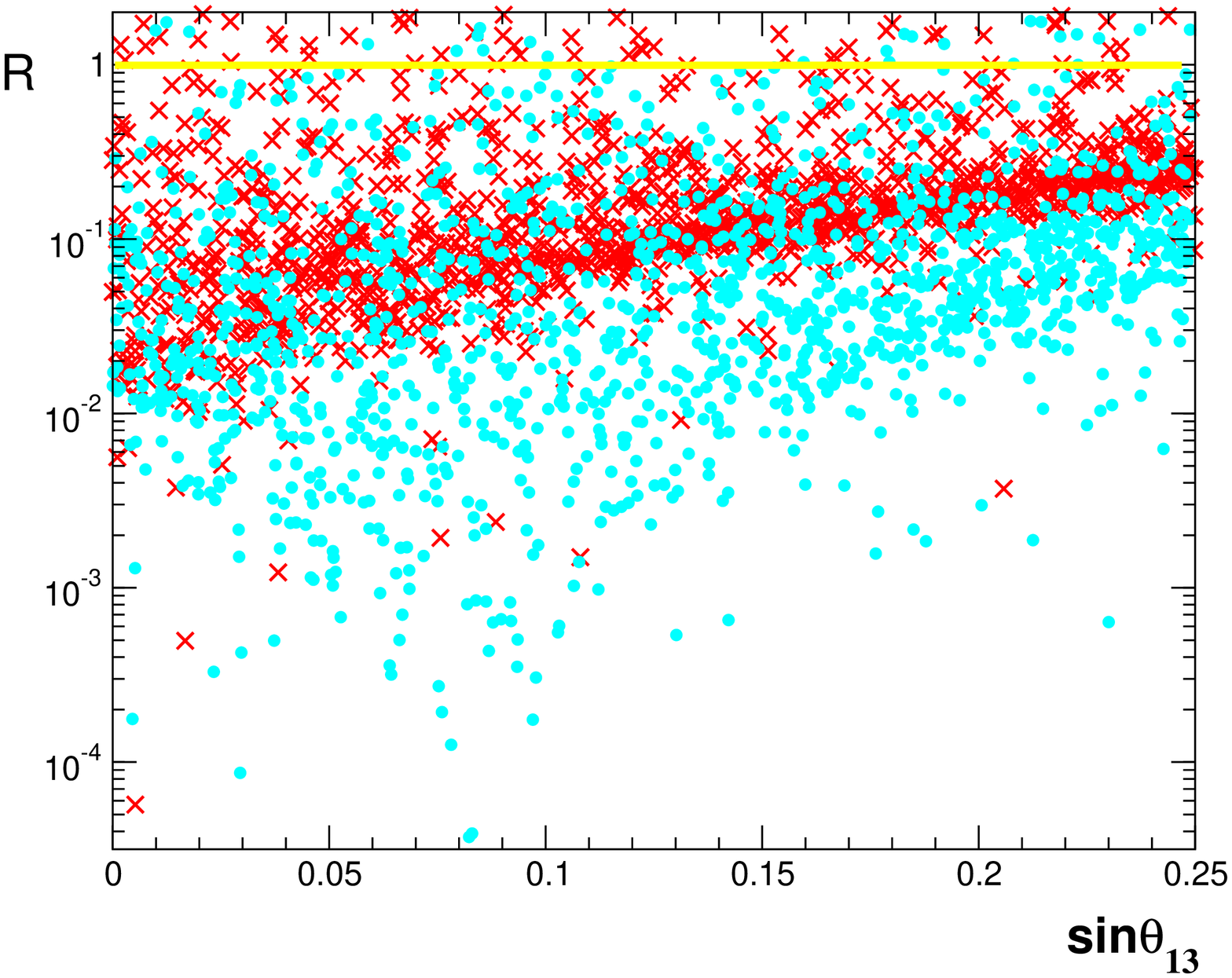}\\
\end{center}
\vspace{0.5 cm}
\includegraphics[angle=270,width=7cm]{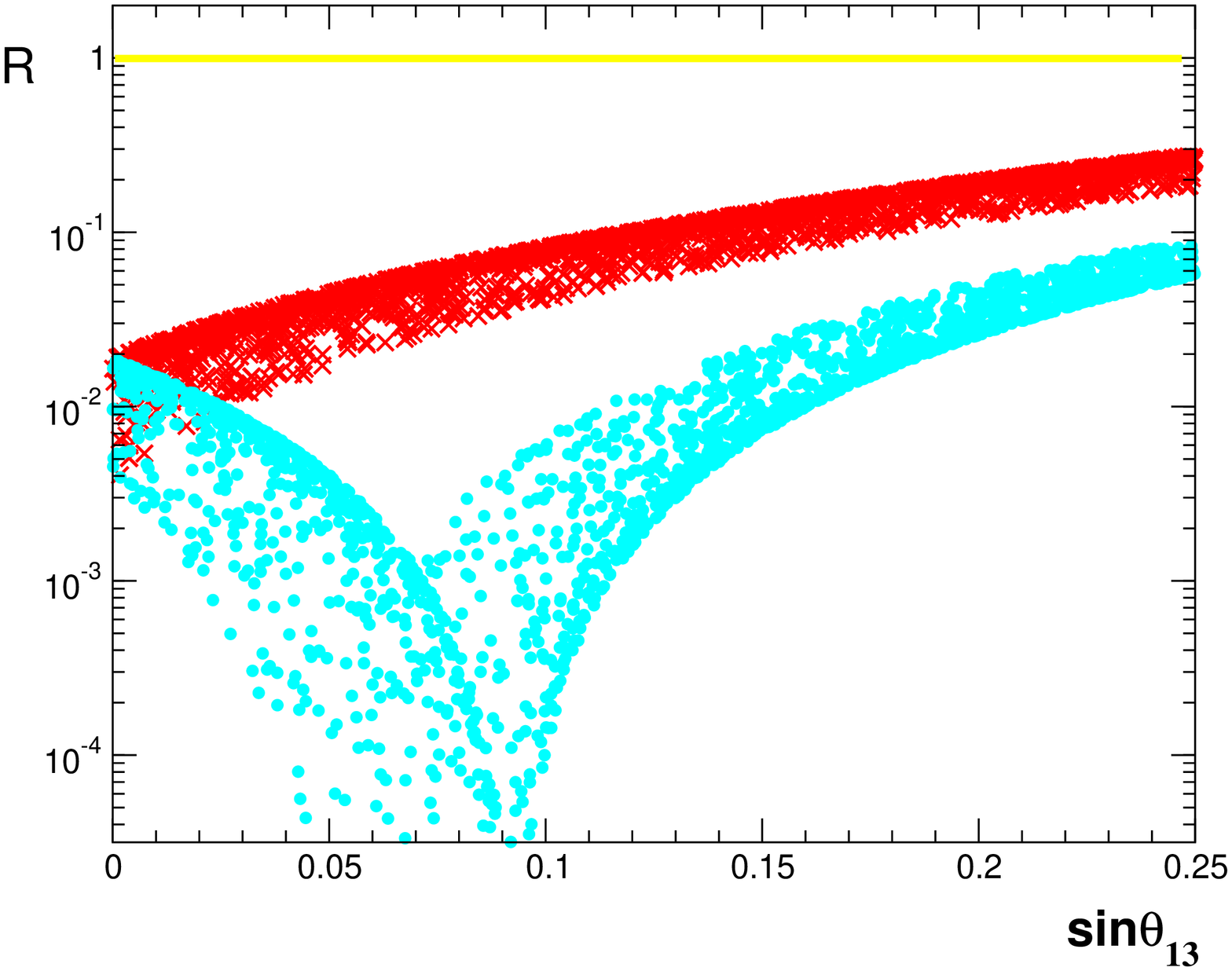}
\hspace{0.5 cm}
\includegraphics[angle=270,width=7cm]{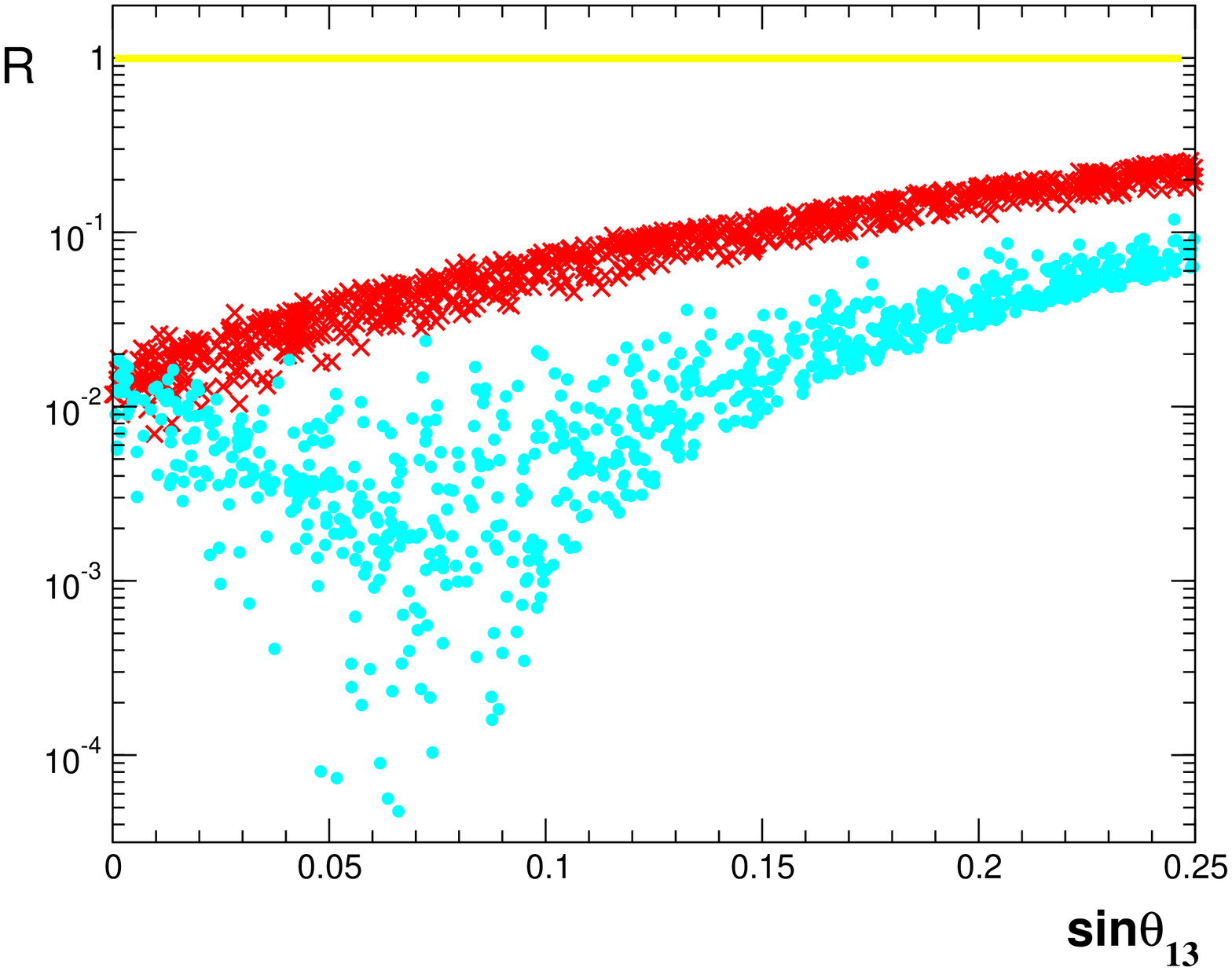}
\vspace{0.5 cm} \begin{center} \caption{\label{fig:three} $R =
\BR(\mu\to e\gamma)/\BR(\tau\to \mu\gamma)$ for different ranges of
the CPV parameters. The light blue circles correspond to $\delta =
0$, the red crosses to $\delta = \pi$. {\em Upper panel}: general
result with CPV phases after imposing the leptogenesis constraint.
{\em Left panel}: no leptogenesis CPV phases. {\em Lower panel}: CPV
case after imposing the leptogenesis constraint, with $M_\nu >
10^{15}$~GeV and $\phi_i < 0.1$. In these plots, $c
\in [0.001,1]$ and $m_\nu^{min} \in
[10^{-4},0.6]$~eV. }
\end{center}
\end{figure}

In the lower panel of Fig.~\ref{fig:four} we plot the $\mu\rightarrow
e \gamma$ branching ratio as a function of $\Lambda_{\rm LFV}$.  We
generated a large number of events extracting randomly all the
parameters in a wide range (see figure caption) and
evaluated $\cB(\mu\rightarrow e \gamma)$ only for those events
yielding the baryon asymmetry within the observed range.\footnote{~Having explicitly 
checked that  the Majorana phases of $\upmns$ have a negligible impact on the 
$l_i\rightarrow l_j\gamma$ rates, these phases have been set to zero both in Fig.~\ref{fig:four} 
and in Fig.~\ref{fig:three}.}
Despite the spread of the points --due to the large number of parameters
involved--
a clear correlation between the values of the branching
ratio and $\Lambda_{\rm LFV}$ emerges. A similar correlation holds also for $\cB(\tau \to \mu \gamma)$
and for other $\mu \to e$ transitions, that can be expressed in
terms of $\cB(\mu \to e \gamma)$ up to known phase space integrals and
unknown ratios of Wilson coefficients~\cite{Cirigliano:2006su}.
This allows us
to identify the ranges of $\Lambda_{\rm LFV}$ that current and future
low-energy LFV experiments can probe. In particular, it is worth stressing that
 $\mu\rightarrow e\gamma$ is already probing new-physics
scales higher than those probed by FCNC transitions in the quark
sector \cite{MFV_quark}.  If no signal is observed by the MEG
experiment at PSI, scales as high as 100 TeV will be excluded, in
contradiction with the natural expectation of the MLFV
hypothesis: therefore a definite prediction of this framework is that
$\mu\rightarrow e \gamma$ should be observed soon.

The second information that can be extracted from the leptogenesis
results is the allowed range of the CP violating parameters $\phi_i$
appearing in the matrix $H$ in Eq.~(\ref{eq:phi}).
The impact of these parameters in the predictions of the 
$\BR(\mu\rightarrow e \gamma)/\BR(\tau\rightarrow\mu\gamma)$ 
ratio is illustrated in Fig.~\ref{fig:three}.
In the upper panel  we plot 
$\BR(\mu\rightarrow e \gamma)/\BR(\tau\rightarrow\mu\gamma)$ as a
function of the $s_{13}$ parameter of the $\upmns$ matrix.
We extracted randomly all the
parameters contained in $\lambda_\nu$ as in Fig.~\ref{fig:four}, 
selecting the events satisfying the leptogenesis constraint. The same
quantity is plotted in the left panel of Fig.~\ref{fig:three} with the
leptogenesis phases set to zero. As can be seen, in general the
inclusion of the high-energy phases largely spoils the CP-conserving
predictions; however, the pattern
$\BR(\mu\rightarrow e \gamma) < \BR(\tau\rightarrow\mu\gamma)$
is still preserved. Indeed the deviations from the CP conserving case
are almost completely driven by the increase of $\cB(\mu\rightarrow e\gamma)$,
while $\cB(\tau\rightarrow \mu \gamma)$ is only marginally affected.
This fact can be easily understood:
in the CP-conserving limit $(\Delta_{\rm FCNC})_{\mu e}$ is suppressed
by the small values of $s_{13}$ and $\Delta m_\nu^{\rm sol}$~\cite{Cirigliano:2005ck}.
Even if not large in magnitude, the flavour off-diagonal CPV parameters of $H$
can spoil these suppression factors
yielding to larger values of $\BR(\mu\rightarrow e \gamma)$.
Their effect is much smaller for $\cB(\tau \rightarrow \mu \gamma)$ since
$(\Delta_{\rm FCNC})_{\tau \mu}$ is proportional to the largest
neutrino mass splitting ($\Delta m_\nu^{\rm atm}$)
already in the CP-conserving case.
 
The strong dependence of FCNC amplitudes on the complex part of
the matrix $R$ defined in Eq.~(\ref{eq:R}) is an intrinsic property of
the see-saw mechanism with quasi-degenerate right-handed neutrinos,
and is not specific of the MLFV framework \cite{Petcov2}.
As far as MLFV models are concerned, we find that there is an interesting 
regime of small phases where we recover the strong predictive power 
of the CP-conserving case (see Fig.~\ref{fig:three} right panel).
This regime holds for high values of $M_\nu$
($M_\nu \sim 10^{15}$ GeV),
as shown in the left panel of Fig.~\ref{fig:two}. 
Such high values are quite welcome
in grand-unified models and give rise to a 
natural order of magnitude for $\lambda_\nu$ (with maximal eigenvalue
of order one, in close analogy with the top-quark Yukawa coupling).
On the other hand, high values of $M_\nu$ enhance 
the tension with the lower bound on $\Lambda_{\rm LFV}$ 
set by  $\BR(\mu\rightarrow e \gamma)$.
Besides these theoretical considerations, 
this regime is particularly interesting from a 
pure phenomenological point of view:
the comparison of the precise predictions of FCNC ratios
with experimental data provides a powerful testing tool.

\section{Conclusions}
\label{sect:conclusions}

Within the quark sector the MFV hypothesis provides a natural
solution to the flavour problem: it is quite impressive that
the huge amount of experimental data available on quark-flavour
mixing is compatible with the hypothesis that the
Yukawa couplings are the only sources
of flavour symmetry breaking \cite{MFV_quark}.
If there is a deep dynamical reason behind this phenomenological
observation, it is natural to expect a similar mechanism at work
--at least to some extent-- also in the lepton sector.

In this work we have analysed the phenomenon of CP
violation in the generic class of models with three
right-handed neutrinos satisfying the criterion of
MLFV (with extended field content) proposed
in Ref.~\cite{Cirigliano:2005ck}: a global
lepton-flavour symmetry $SU(3)_{L_L}\times SU(3)_{e_R} \times O(3)_{\nu_R}$
broken only by two irreducible Yukawa
structures $\lambda_e \sim (\bar 3,3,1)$
and $\lambda_\nu \sim (\bar 3,1,3)$. 
Explicit realizations of this general scenario can be 
implemented in the minimal supersymmetric SM (MSSM) 
with see-saw mechanism (see e.g.~Ref.~\cite{Borzumati}), 
or in models with TeV-scale neutrinos, 
such as those analysed in Ref.~\cite{Hambye:2004jf,resonance}. 
We have shown that in this class of models 
it is possible to generate a phenomenologically
acceptable leptogenesis, satisfying the MLFV hypothesis, 
only if the overall scale of right-handed neutrinos 
is sufficiently high: $M_\nu \gsim 10^{12}$~GeV.

This result was not trivial a priori given the
strong restrictions on the model imposed by the 
flavour symmetry, which in first approximation 
implies degenerate right-handed neutrinos. 
By means of a general analysis of the CPV invariants,
we have demonstrated that the theory possesses
enough physical CPV parameters contributing
to leptogenesis, even in the limit $\lambda_e \to 0$. 
As far as the parameterization of $\lambda_\nu$ is concerned, 
it turns out that in this class of models $\lambda_\nu$
can be unambiguously expressed in terms of 9 low-energy parameters
(3 light-neutrino mass eigenvalues and 6 parameters of $\upmns$),
one overall scale ($M_\nu$),
and only 3 high-energy CPV parameters (the remnants of the
matrix elements of the Casas-Ibarra matrix $R$~\cite{Casas:2001sr},
after imposing the $O(3)_{\nu_R}$ invariance). The latter
are responsible for leptogenesis.

From a dynamical point of view, the mechanism which 
allows leptogenesis in this framework is the breaking 
of the exact $\nu_R$ degeneracy by means 
of Yukawa-induced corrections. This mechanism 
is necessarily present in any underlying theory 
because of radiative corrections. The MLFV hypothesis
forces it to be the only source of breaking of the
$\nu_R$ degeneracy. This well-defined pattern 
for the breaking of the $\nu_R$ degeneracy is the 
origin of the interesting constraints on $M_\nu$ 
derived in this framework and, more generally, 
of the good predictive power of this class 
of models. We stress that this appealing feature 
is not present in generic models with 
quasi-degenerate heavy neutrinos.
The $M_\nu \gsim 10^{12}$~GeV bound we have derived 
is the result of a numerical scan with approximate 
solutions to the Boltzmann equations, 
therefore it cannot be considered as a strict lower limit. 
Rather, it should be regarder as the natural lower 
scale for successful leptogenesis in this generic framework. 
In the appendix we have shown how a semi-quantitative understanding 
of this lower limit can be obtained in terms of simple 
analytical formulae.
 
The consequences on low-energy FCNC rates
following from the requirement of successful leptogenesis
have been investigated, with particular attention to
the $\mu\rightarrow e\gamma$ and $\tau\rightarrow \mu\gamma$
processes. The most striking consequence of this
additional requirement is the lower bound on $M_\nu$,
which breaks the ambiguity in the normalization of
$\lambda_\nu$ and thus of the
FCNC rates. 
As a result of the lower bound on $M_\nu$,
for natural values of $\Lambda_{\rm LFV}$ in the TeV 
range (i.e.~assuming new particles carrying lepton flavour 
at the TeV scale, such as for instance within the MSSM), 
the $\mu\rightarrow e\gamma$ rate turns out
to be quite close to its present exclusion limit, and
well within the reach of the MEG experiment. For the same reason, 
the rates for $\mu \to e \bar{e} e$ and 
and $\mu \to e$ conversion in nuclei are expected to be 
one or two orders of magnitude below the present experimental limits.

On general grounds, if the three high-energy CPV parameters are
not vanishing the predictions for low-energy FCNC
transitions derived in Ref.~\cite{Cirigliano:2005ck}
in the CP-conserving limit can be substantially modified.
After imposing the leptogenesis conditions, we find that
$\cB(\mu\rightarrow e\gamma)$ can easily be enhanced with
respect to the CP-conserving case, while
$\cB(\tau\rightarrow \mu\gamma)$ is quite stable. As a result,
in general the specific pattern  $\BR(\mu\rightarrow e\gamma) \ll
\BR(\tau\rightarrow \mu\gamma)$ \cite{Cirigliano:2005ck}
does not hold, but only the weaker condition  $\BR(\mu\rightarrow e\gamma)<
\BR(\tau\rightarrow \mu\gamma)$ is satified.
Interestingly, we have also found that there
exists a region of the parameter space where
leptogenesis condition can be accomplished
with very small CPV parameters, such that
the full predictive pattern of FCNC transitions
holds as in the CP-conserving case. 

\section*{Acknowledgment}
We thank Thomas Hambye for pointing our attention
on Ref.~\cite{Hambye:2004jf} and for useful
discussions. We also thank Ben Grinstein, Serguey
Petcov  and Mark Wise  for useful comments.
The work of VC was carried out under the auspices of the National Nuclear Security Administration 
of the U.S. Department of Energy at Los Alamos National Laboratory under Contract No. DE-AC52-06NA25396.

\section*{Note Added}
During the refereeing process of this article, a preprint addressing 
the viability of leptogenesis in the MLFV framework has appeared \cite{BurasBranco}. 
The analysis of Ref.~\cite{BurasBranco} confirms our main result about the 
viability of leptogenesis in MLFV via Yukawa-induced corrections, 
and all our analytical treatment for the inclusion of CPV effects 
in this framework.
As far as the estimate of $\eta_B$ is concerend, 
the qualitative behavior with almost  
no dependence for $M_\nu \lsim 10^{11}$~GeV and linear growth 
for $M_\nu \gsim 10^{11}$~GeV is also confirmed. However, the authors 
of Ref.~\cite{BurasBranco} claim that a more refined treatment 
of Boltzmann equations with respect to 
Ref.~\cite{DiBari} (i.e.~the approach we have adpted) 
and the inclusion of flavour-dependent effects, raises the maximal values of 
$\eta_B(M_\nu)$ slightly above the $10^{-10}$ level also 
in the low-$M_\nu$ region.  
If confirmed, this result would weaken our conclusion about the 
lower bound on $M_\nu$. 
However, we stress that even in the approach of Ref.~\cite{BurasBranco}
only a marginal region of the parameter space can account
for the experimental value of $\eta_B$ for $M_\nu\lesssim 10^{12}$ GeV.
We still expect that the conclusions reached in the present 
work by means of a less sophisticated analysis of leptogenesis  
represent the basic features expected in most models with MLFV. 

\appendix

\section{On the scaling of $\eta_B$ with $M_\nu$} 
We present here a simplified discussion to illustrate  
the scaling of $\eta_B$ with $M_\nu$ (and the resulting lower 
bound on $M_\nu$) in a simple analytical form. In particular, 
we prove that in the high-$M_\nu$ regime (where $\lambda_e$ terms 
can be neglected) $\eta_B = \kappa  M_\nu$ and 
we provide a reasonable estimate of  the dimensional parameter $\kappa$. 

Let us start from the formula for the baryon asymmetry:
\beq
\eta_B = 9.6 \cdot 10^{-3} \times d(K_1 + K_2  + K_3) \times (\epsilon_1 + \epsilon_2 + \epsilon
_3)~.
\eeq
The dilution factor $ d(K_1 + K_2  + K_3)$ does not depend on $M_\nu$. 
The $M_\nu$ dependence comes from the asymmetries $\epsilon_i$,  which near 
resonance are described by Eq.~(\ref{eq:simpl}). In order 
to identify the scaling of the $\epsilon_i$ with $M_\nu$, we count 
the powers of $\lambda_\nu$ insertions and use the relation
 $\lambda_\nu \sim M_\nu^{1/2}$ (see Eq.~(\ref{eq:R})).

The denominator in Eq.~(\ref{eq:simpl}) scale as $\lambda_\nu^4 \propto  M_\nu^2$. 
The analysis of the imaginary part in the numerator is slightly more
involved. In general 
\beq
\bar{\lambda}_\nu\bar{\lambda}_\nu^\dagger = {\bar U} \, \lambda_\nu \lambda_\nu^\dagger \, 
{\bar U}^\dagger~,
\eeq
where $\bar U$ is the unitary matrix which diagonalizes the right-handed mass matrix: 
\beq
\bar U^T \ M_R \ \bar U = M_R^{\rm diag}~.
\eeq
Focusing on a specific structure of $M_R$  with only  $c_{11}$ and $c_{21}$ non-zero,  
we have:
\beq
M_\nu^{-1} M_R = I + c_{11} \delta_1 + c_{21} \delta_2 \qquad \delta_1 = h_\nu + h_\nu^T  \qquad 
\delta_{2} = h_\nu^2 + (h_\nu^T)^2 
\eeq
where, as usual, $h_\nu=\lambda_\nu \lambda_\nu^\dagger$.
Let us also assume the perturbative behavior  $c_{11}\sim c$ and $c_{21}\sim c^2$ with $c<1$,
which allows us to perform a perturbative diagonalization of $M_R$. In general, we can write  
$\bar U = O_1 \, O_2$, where $O_{1,2}$ are real-orthogonal matrices and  $O_1$  
is the matrix diagonalizing  $\delta_1$:
\beq 
M_\nu^{-1}~ O_1^T M_R O_1 
= \bar{M} + c_{21}  \bar{\delta}_2~,     \qquad  \bar{\delta}_2  = O_1^T  \delta_2 O_1
\eeq
(with $\bar{M}$ diagonal). As can be explicitly verified in perturbation 
theory, $O_1 \sim O(1)$ and in first approximation does not depend on $M_\nu$. 
Now we can proceed with the diagonalization of $O_1^T M_R O_1$ by means of 
$O_2$. To first non-trivial order in $c_{22} \bar{\delta_2}$ we can write 
$O_2 \approx I + A$, with
\beq
A^{ij} = \frac{c_{21}  \bar{\delta}_{2}^{ij}}{ \bar{M}^{jj} - \bar{M}^{ii}}  \propto  \frac{h_\nu^2}{h_\nu}  
 \sim \lambda_\nu^2 \propto  M_\nu~.
\eeq
Given the quadratic dependence on $h_\nu$ in the numerator, in this case we get a 
non-trivial scaling with $M_\nu$. Putting together the above results and recalling 
that 
\beq
{\rm Im}  \left[  (O_1 \, \lambda_\nu \lambda_\nu^\dagger \, O_1^T)_{ij} \ 
(O_1 \, \lambda_\nu \lambda_\nu^\dagger \, O_1^T)_{ij} 
\right] = 0
\eeq
(see sect.~\ref{sec:invar}), 
the leading contributions to the asymmetry arise from terms of the type 
\beq
{\rm Im} \left[ \left( \bar{\lambda}_\nu\bar{\lambda}_\nu^\dagger\right)_{ij}^2  \right]   \propto  
{\rm Im}  \left[  (A \,    O_1 \, \lambda_\nu \lambda_\nu^\dagger \, O_1^T)_{ij} \ 
(O_1 \, \lambda_\nu \lambda_\nu^\dagger \, O_1^T)_{ij} 
\right]  \sim    \lambda_\nu^6  \propto  M_\nu^3~. 
\eeq

We are now finally able to evaluate the complete scaling of the baryon 
asymmetry. 
As expected, $\eta_B$ grows linearly with $M_\nu$. More explicitly, we find 
\beq
\eta_B \sim  9.6 \cdot 10^{-3} \times d  \times   \frac{\sqrt{\Delta m^2_{\rm atm}}}{v^2}  \times  
f(\phi_1,\phi_2, \phi_3) 
\times  M_\nu~,
\eeq 
where  $f(\phi_1,\phi_2, \phi_3) \sim   \phi_1 \phi_2 \phi_3 $ up to higher orders in the phases. 
Numerically, using  $d \sim 10^{-3}$ and  $f(\phi_1,\phi_2, \phi_3) \sim O(10^{-1})$  
(moderately large CPV phases) we get
\beq
\eta_B \sim  1.2  \times  10^{-21} \times \frac{M_{\nu}}{\rm GeV}~.
\eeq
Requiring $\eta_B > \eta_B^{\rm exp} = 6.3 \times 10^{-10}$ 
implies $M_\nu > 5 \times 10^{11}$ GeV, in remarkable 
agreement with our findings from the numerical scan.


\begin{thebibliography}{99}
  

\bibitem{Georgi}
R.~S.~Chivukula and H.~Georgi, \plb{188}{1987}{99}.

\bibitem{Hall:1990ac}
  L.~J.~Hall and L.~Randall,
  Phys.\ Rev.\ Lett.\  {\bf 65}, 2939 (1990).

\bibitem{MFV}
G.~D'Ambrosio, G.~F.~Giudice, G.~Isidori and A.~Strumia,
\npb{645}{2002}{155} [hep-ph/0207036].


\bibitem{Cirigliano:2005ck}
V.~Cirigliano, B.~Grinstein, G.~Isidori and M.~B.~Wise,
  Nucl.\ Phys.\ {\bf B 728},  121 (2005) [hep-ph/0507001].

\bibitem{Cirigliano:2006su}
  V.~Cirigliano and B.~Grinstein,
  hep-ph/0601111.

\bibitem{Weinberg:1979sa}
  S.~Weinberg,
  Phys.\ Rev.\ Lett.\  {\bf 43}, 1566 (1979).

\bibitem{Fukugita:1986hr}
  M.~Fukugita and T.~Yanagida,
  Phys.\ Lett.\ B {\bf 174}, 45 (1986).

\bibitem{rossi}
  A.~Rossi,
  Phys.\ Rev.\ D {\bf 66}, 075003 (2002)
  [hep-ph/0207006].


\bibitem{D'Ambrosio:2004fz}
  G.~D'Ambrosio, T.~Hambye, A.~Hektor, M.~Raidal and A.~Rossi,
  Phys.\ Lett.\ B {\bf 604}, 199 (2004)
  [hep-ph/0407312].


\bibitem{Hambye:2005tk}
  T.~Hambye, M.~Raidal and A.~Strumia,
  Phys.\ Lett.\ B {\bf 632}, 667 (2006)
  [hep-ph/0510008].


\bibitem{Hambye:2004jf}
  T.~Hambye, J.~March-Russell and S.~M.~West,
  JHEP {\bf 0407}, 070 (2004)
  [hep-ph/0403183].


\bibitem{MEG}
  M.~Grassi  [MEG Collaboration],
  Nucl.\ Phys.\ Proc.\ Suppl.\  {\bf 149}, 369 (2005)



\bibitem{GonzalezFelipe:2003fi}
  R.~Gonzalez Felipe, F.~R.~Joaquim and B.~M.~Nobre,
  Phys.\ Rev.\ D {\bf 70} (2004) 085009
  [hep-ph/0311029].
  G.~C.~Branco, R.~Gonzalez Felipe, F.~R.~Joaquim and B.~M.~Nobre,
  Phys.\ Lett.\ B {\bf 633} (2006) 336
  [hep-ph/0507092].


\bibitem{CPasymmetries}
  L.~Covi, E.~Roulet and F.~Vissani,
  Phys.\ Lett.\ B {\bf 384}, 169 (1996)
  [hep-ph/9605319];
  M.~Flanz, E.~A.~Paschos, U.~Sarkar and J.~Weiss,
  Phys.\ Lett.\ B {\bf 389}, 693 (1996)
  [hep-ph/9607310];
  A.~Pilaftsis,
  Phys.\ Rev.\ D {\bf 56}, 5431 (1997)
  [hep-ph/9707235];
  W.~Buchmuller and M.~Plumacher,
  Phys.\ Lett.\ B {\bf 431}, 354 (1998)
  [hep-ph/9710460];
  G.~C.~Branco, R.~Gonzalez Felipe, F.~R.~Joaquim and M.~N.~Rebelo,
  Nucl.\ Phys.\ B {\bf 640} (2002) 202
  [hep-ph/0202030];
  G.~C.~Branco {\em et al.} 
  Phys.\ Rev.\ D {\bf 67} (2003) 073025
  [hep-ph/0211001].

\bibitem{Ellis}
  J.~R.~Ellis, J.~Hisano, S.~Lola and M.~Raidal,
  Nucl.\ Phys.\ B {\bf 621}, 208 (2002)
  [hep-ph/0109125].


\bibitem{Lepto2}
  T.~Hambye, Y.~Lin, A.~Notari, M.~Papucci and A.~Strumia,
  Nucl.\ Phys.\ B {\bf 695}, 169 (2004)
  [hep-ph/0312203];
  A.~Anisimov, A.~Broncano and M.~Plumacher,
  Nucl.\ Phys.\ B {\bf 737}, 176 (2006)
  [hep-ph/0511248];
  A.~Strumia and F.~Vissani,
  hep-ph/0606054.


\bibitem{Santamaria:1993ah}
  A.~Santamaria,
  Phys.\ Lett.\ B {\bf 305}, 90 (1993)
  [hep-ph/9302301].

\bibitem{Bernabeu:1986fc}
  J.~Bernabeu, G.~C.~Branco and M.~Gronau,
  Phys.\ Lett.\ B {\bf 169}, 243 (1986).

\bibitem{Branco:1989bn}
  G.~C.~Branco, M.~N.~Rebelo and J.~W.~F.~Valle,
  Phys.\ Lett.\ B {\bf 225}, 385 (1989).

\bibitem{Branco:2001pq}
  G.~C.~Branco, T.~Morozumi, B.~M.~Nobre and M.~N.~Rebelo,
  Nucl.\ Phys.\ B {\bf 617}, 475 (2001)
  [hep-ph/0107164].



\bibitem{Casas:2001sr}
  J.~A.~Casas and A.~Ibarra,
  Nucl.\ Phys.\ B {\bf 618}, 171 (2001)
  [hep-ph/0103065].


\bibitem{Pascoli:2003rq}
  S.~Pascoli, S.~T.~Petcov and C.~E.~Yaguna,
  Phys.\ Lett.\ B {\bf 564}, 241 (2003)
  [hep-ph/0301095].




\bibitem{DiBari}
  S.~Blanchet and P.~Di Bari,
  hep-ph/0603107.

\bibitem{resonance}
  J.~R.~Ellis, M.~Raidal and T.~Yanagida,
  Phys.\ Lett.\ B {\bf 546}, 228 (2002) 
  [hep-ph/0206300];
  A.~Pilaftsis and T.~E.~J.~Underwood,
  Nucl.\ Phys.\ B {\bf 692}, 303 (2004)
  [hep-ph/0309342].

\bibitem{flavour}
  R.~Barbieri, P.~Creminelli, N.~Tetradis, A.~Strumia,
  \npb{575}{2000}{61} [hep-ph/9911315];
  A.~Pilaftsis and T.~E.~J.~Underwood,
  Phys.\ Rev.\ D {\bf 72}, 113001 (2005)
  [hep-ph/0506107];
  A.~Abada {\em et al.}
  JCAP {\bf 0604}, 004 (2006)
  [hep-ph/0601083];
  E.~Nardi, Y.~Nir, E.~Roulet and J.~Racker,
  JHEP {\bf 0601}, 164 (2006)
  [hep-ph/0601084].


\bibitem{FCNC_exp_rev}
See e.g.~T.~Mori,
hep-ex/0605116.


\bibitem{Petcov2}
  S.~T.~Petcov and T.~Shindou,
  hep-ph/0605151.


\bibitem{MFV_quark}
A.~J.~Buras {\em et al.}, \plb{500}{2001}{161}
[hep-ph/0007085];
M.~Bona {\it et al.}
hep-ph/0605213.

\bibitem{Borzumati}
F.~Borzumati and A.~Masiero,
\prl{57}{1986}{961}.

\bibitem{BurasBranco}
  G.C.~Branco {\em et al.}, 
  hep-ph/0609067.

\end{thebibliography}
\end{document}